\documentclass[sigconf]{acmart}

\usepackage{array} 
\usepackage[utf8]{inputenc}
\usepackage{graphicx}
\usepackage{listliketab}
\usepackage{enumitem}
\usepackage{soul} 
\sethlcolor{yellow}
\usepackage{booktabs}
\usepackage{colortbl}
\usepackage{url}
\usepackage[hyphens]{xurl} 
\hypersetup{breaklinks=true}

\copyrightyear{2025}
\acmYear{2025}
\setcopyright{cc}
\setcctype{by}
\acmConference[IUI '25]{30th International Conference on Intelligent User Interfaces}{March 24--27, 2025}{Cagliari, Italy}
\acmBooktitle{30th International Conference on Intelligent User Interfaces (IUI '25), March 24--27, 2025, Cagliari, Italy}\acmDOI{10.1145/3708359.3712082}
\acmISBN{979-8-4007-1306-4/25/03}

\begin{document}
\title{ArtInsight: Enabling AI-Powered Artwork Engagement for\\ Mixed Visual-Ability Families}

\author{Arnavi Chheda-Kothary}
\orcid{0000-0001-8627-0412}
\affiliation{\institution{University of Washington}
\city{Seattle}
\country{USA}}
\email{chheda@cs.washington.edu}

\author{Ritesh Kanchi}
\orcid{0009-0006-7978-0821}
\affiliation{\institution{University of Washington}
\city{Seattle}
\country{USA}}
\email{rkanchi@uw.edu}

\author{Chris Sanders}
\orcid{0009-0002-1380-7025}
\affiliation{\institution{University of Washington}
\city{Seattle}
\country{USA}}
\email{ck.sanders043@gmail.com}

\author{Kevin Xiao}
\orcid{0009-0009-8857-1046}
\affiliation{\institution{University of Washington}
\city{Seattle}
\country{USA}}
\email{xckevin@cs.washington.edu}

\author{Aditya Sengupta}
\orcid{0009-0008-1956-5331}
\authornote{Author contributed as a high school intern at the University of Washington.}
\affiliation{\institution{University of Washington}
\city{Seattle}
\country{USA}}
\affiliation{\institution{The Overlake School}
\city{Bellevue}
\country{USA}}
\email{adityasngpta@gmail.com}

\author{Melanie Kneitmix}
\orcid{0009-0001-9856-1038}
\affiliation{\institution{University of Washington}
\city{Redmond}
\country{USA}}
\email{mekne@cs.washington.edu}

\author{Jacob O. Wobbrock}
\orcid{0000-0003-3675-5491}
\affiliation{\institution{University of Washington}
\city{Seattle}
\country{USA}}
\email{wobbrock@uw.edu}

\author{Jon E. Froehlich}
\orcid{0000-0001-8291-3353}
\affiliation{\institution{University of Washington}
\city{Seattle}
\country{USA}}
\email{jonf@cs.uw.edu}

\renewcommand{\shortauthors}{Chheda-Kothary et al.}

\begin{abstract}
We introduce \textit{ArtInsight}, a novel AI-powered system to facilitate deeper engagement with child-created artwork in mixed visual-ability families. ArtInsight leverages large language models (LLMs) to craft a respectful and thorough initial description of a child's artwork, and provides: creative AI-generated descriptions for a vivid overview, audio recording to capture the child's own description of their artwork, and a set of AI-generated questions to facilitate discussion between blind or low-vision (BLV) family members and their children. Alongside ArtInsight, we also contribute a new rubric to score AI-generated descriptions of child-created artwork and an assessment of state-of-the-art LLMs. We evaluated ArtInsight with five groups of BLV family members and their children, and as a case study with one BLV child therapist. Our findings highlight a preference for ArtInsight's longer, artistically-tailored descriptions over those generated by existing BLV AI tools. Participants highlighted the creative description and audio recording components as most beneficial, with the former helping \textit{``bring a picture to life''} and the latter centering the child's narrative to generate context-aware AI responses. Our findings reveal different ways that AI can be used to support art engagement, including before, during, and after interaction with the child artist, as well as expectations that BLV adults and their sighted children have about AI-powered tools.

\end{abstract}
\keywords{Accessibility, blind or low-vision, mixed-ability families, children's artwork, AI}

\begin{teaserfigure}
\centering
\includegraphics[width=0.9\textwidth]{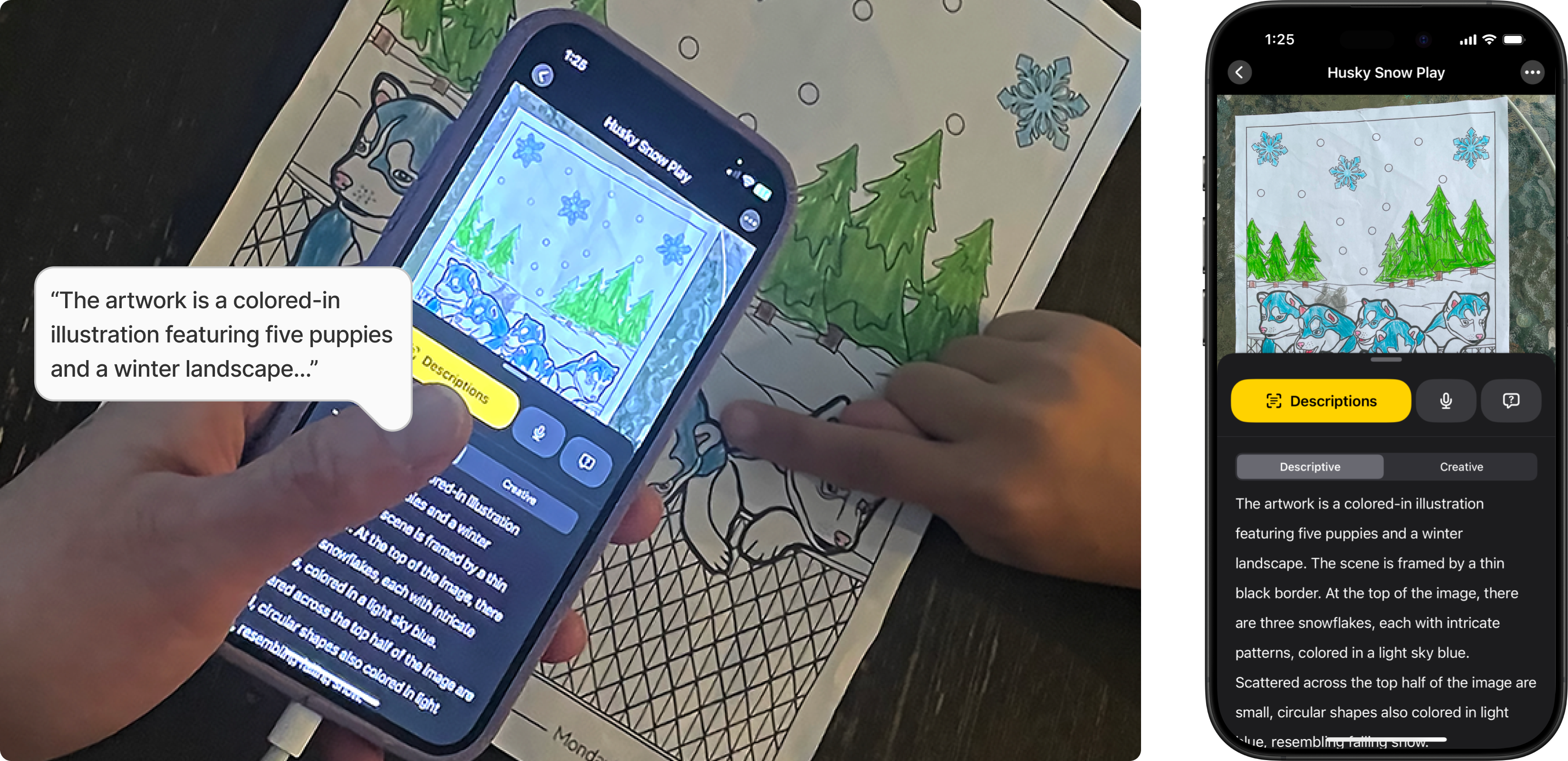}
\Description{This image consists of two panes. On the left: A photo of a grandmother's hand holding an iPhone running the ArtInsight application on it as her granddaughter's hand (to the right) points out specific things in her artwork. The art is done on a page of a coloring book with several husky puppies set against a snowy forest background. The ArtInsight application is analyzing the granddaughter's artwork. On the right: a close-up of the ArtInsight application running on an iPhone, describing the husky puppy artwork from the photo.}
\caption{ArtInsight is a novel AI-powered mobile prototype that describes child-created artwork and facilitates discussion in mixed visual-ability families. Left: A blind grandmother uses ArtInsight to explore her sighted granddaughter's artwork as her granddaughter simultaneously explains different parts of her art. Right: The ArtInsight system, describing the artwork.}
\label{teaser}
\end{teaserfigure}

\maketitle
\section{Introduction}

\begin{quote}
    \textit{[My son] would describe [his art] to me as best as he could... I'd go, `Okay, that's a good drawing.' But I was too dry. I was trying to acknowledge his drawing, but I really couldn't compliment him because I didn't really know what was there... you want your child to feel like you're acknowledging their work, but in reality... you feel defeated. [With ArtInsight], I feel more truly connected to the drawing.} --- \textbf{P4}
\end{quote}

Creating and interacting with art is invaluable for children’s developmental and social milestones \cite{importance_for_early_development_2024, Lynch_2012, Rymanowicz_2021} as well as a critical engagement opportunity for parents and family members \cite{how_parents_can_help_2024}; however, given its intrinsically visual nature, accessing a child's art and art process can be more challenging for blind or low-vision (BLV) family members. Recent work by \citet{chhedakothary2024} demonstrates how advances in Artificial Intelligence (AI) may help BLV family members more deeply engage with their children's artwork, but highlights the lack of specific tools for child-created artwork interpretation. While BLV individuals are increasingly employing AI to understand visual artifacts such as videos \cite{Bodi_Video, Yuksel_Video_HumanInTheLoop}, images \cite{Huh_GenAssist}, and scene descriptions \cite{gonzalez_CHI, be_my_ai_2024, seeing_ai}, AI-powered interpretations of child-created artwork need to be more respectful, less presumptive, and more detailed than existing BLV AI tools, all while centering the child's narrative \cite{chhedakothary2024}.

We introduce \textit{ArtInsight}, a novel AI-powered mobile prototype that recognizes and describes features of child-created artwork and facilitates discussion between BLV family members and their children. Informed by work in mixed-ability family interactions \cite{cuddlingup, chhedakothary2024}, promoting parent-child dialogue \cite{Dietz_ContextQ, Zhang_StoryBuddy}, and AI design considerations \cite{GAI_DesignPrinciples, bennett_itscomplicated, darth_vader, Kim_DescriptionsForComics, Das_ProvenanceToAberrations}, ArtInsight is composed of two high-level parts: (1) a custom prompted GPT-backed AI agent that analyzes and describes child-created artwork; and (2) three complementary components aimed at centering human agency and facilitating discussion, including the ability to control the “personality” of the AI-generated descriptions (more "creative" \textit{vs.} "descriptive"), capturing audio recordings of children’s artwork descriptions, and AI-generated questions about the artwork to provoke family member-child dialogue. Our overarching goal is not to replace interactions between BLV family members and their children, but rather to augment these interactions through Human-AI capabilities. For example, using ArtInsight, a grandmother can take a picture of her granddaughter’s artwork from a coloring book (Figure \ref{teaser}) to receive an initial AI-generated description and receive questions to guide their conversation. Alternatively, the grandmother can record her granddaughter’s description of the artwork while they are together, allowing the system to update the AI-generated descriptions with the child's own context as well as enabling the grandmother to preserve her granddaughter's audio description.

We evaluated ArtInsight at two layers: the AI backend layer and the user interaction layer \textit{i.e.,} the end-to-end system. To determine the best AI backend, we conducted an experiment across four state-of-the-art models: Claude 3.5 Sonnet \cite{Claude35}, GPT-4 Turbo \cite{GPT-4Turbo}, GPT-4o \cite{GPT-4o}, and Gemini 1.5 Flash \cite{Gemini}. Because existing metrics were insufficient for evaluating AI-generated descriptions of child artwork, we also introduce a custom scoring rubric to evaluate the level of detail, assumptions made, and reductive language used in descriptions of child-created artwork. GPT-4o outperformed the other models, delivering highly detailed and respectful descriptions. 

To evaluate the end-to-end ArtInsight system, we conducted a user study with five groups of BLV family members and their children as well as a qualitative case study with a blind family therapist who actively uses art in her sessions with children. For the first study, we conducted both comparison and exploratory tasks with the mixed visual-ability family groups. We first asked participants to compare ArtInsight descriptions of child-created artwork to descriptions generated by \textit{Be My AI} \cite{be_my_ai_2024}, a popular vision application for BLV people. We then evaluated the three additional components specific to ArtInsight (the "creative" description, audio recording and AI re-prompting, and AI-generated questions) through open-ended tasks and semi-structured interviews. In the case study with the blind family therapist, T1, we began with asking about her experiences and challenges using artwork in therapy sessions with children. T1 explored ArtInsight in a similar fashion to the first study, \textit{i.e.,} with comparison and exploration tasks, but we focused the interview questions on the potential benefits of ArtInsight for T1’s work as a family therapist to gain broader insights into AI-powered systems for interpreting children’s artwork.

Findings from our first study reveal that BLV family members and their children find ArtInsight's descriptions of artwork more \textit{useful} than Be My AI's descriptions. Of the additional ArtInsight components, BLV family members and their children highlighted the creative descriptions and the audio recording with AI re-prompting as their favorite. Participants expressed primarily wanting to use ArtInsight \textit{before} or \textit{during} artwork-focused interactions with their child, though one BLV mother expressed wanting to share the AI descriptions of her child's art \textit{post hoc} with her friends as alternative text on social media. Our findings also revealed differences in expectations from AI across BLV family members and their children, such as tolerance for inaccuracies.

Our case study highlights how ArtInsight could enable BLV therapists to independently conduct therapy sessions with children, in which artwork plays a key role. The participant therapist also reflected on constraints of using an AI artwork interpretation system during child therapy sessions, such as limited time, wanting to avoid technology in the sessions, and protecting the privacy of child patients. The participant's evaluation of ArtInsight showed a different usage of the audio recording component, which she used to capture an interaction that included herself asking questions as opposed to only a child's description.

In summary, our work contributes: (1) an AI-powered artwork understanding system, \textit{ArtInsight}, to enable deeper artwork engagement between BLV family members and their children, (2) findings from an empirical evaluation of ArtInsight, and (3) a novel scoring rubric to guide AI descriptions of children's artwork.

\section{Background and Related Work}
Our work builds on prior research in AI-based technology to promote family interactions, AI tools for BLV people to enable access to visual content, and Human-AI systems that enable corrections and context augmentations to AI output. 

\subsection{AI in Family Interactions}
An emerging body of work explores the role of AI in family interactions, specifically to augment interactions between parents and children. AI-supported co-reading scenarios are a prominent theme \cite{Zhang_StoryBuddy, Dietz_ContextQ, Lin_FishScales, Xu_Bilingual}. \citet{Dietz_ContextQ} present \textit{ContextQ}, a system supporting parents and children through AI-generated questions to promote dialogue and conversation while co-reading. \citet{Zhang_StoryBuddy}'s \textit{StoryBuddy} similarly presents automated questions throughout the reading of a story, but enables children's interaction with the AI agent both in the presence and in the absence of parents. Additionally, \textit{StoryBuddy} tracks developmental progress through children's responses to these AI-generated questions. \citet{Xu_Bilingual} also employ a bilingual conversational agent to promote language literacy while continuing to foster dialogue and parent-child connections while co-reading. These systems and their study findings inform our design guidelines for ArtInsight---particularly to center the connection between BLV family members and their children while using technology for artwork interpretation.

A smaller body of work investigates the use of AI in mixed-ability family settings. \citet{ASL_DeafChild_HearingParent} leverage AI to enable ASL word retrieval, fostering communication between Hearing parents and their Deaf or Hard-of-Hearing (DHH) children. A related work looks at context-responsive ASL recommendations between Hearing parents and their DHH children \cite{Hossain_ContextResponsiveASL}. \citet{Park_AllAboutPictures_2023} propose smart speakers for BLV parents and their sighted children to co-read picture books. \citet{cuddlingup} discuss researching unobtrusive AI systems that understand both images and text to also support co-reading in mixed-ability families. The researchers also coin the term \textit{``Intimate Assistive Technology''} \cite{cuddlingup}, defined as technology that \textit{``enables individual or collaborative access and fosters interpersonal connection building.''} Our work builds on these past works by expanding the \textit{Intimate Assistive Technology} space to include AI-powered systems that foster increased artwork engagement between BLV family members (\textit{i.e.,} broader than parents) and their sighted children.

\subsection{AI-Powered Tools for BLV People}
Researchers and product teams alike are investigating AI tools for BLV people to navigate both physical and digital spaces. Applications such as \textit{Be My Eyes} \cite{be_my_eyes_2024} and \textit{Seeing AI} now use state-of-the-art AI models for BLV users to take and analyze photos of the physical world around them. \textit{Be My Eyes}, an application experience that is best known for connecting BLV people to human volunteers for sighted assistance, calls this new AI mode \textit{Be My AI} \cite{be_my_ai_2024}.  

One category of research explores \textit{how} BLV people use existing AI systems, such as Be My AI or AI-generated alternative text for images, as well as BLV people's attitudes towards the output of these AI systems \cite{gonzalez_CHI, Das_ProvenanceToAberrations, bennett_itscomplicated, Glazko_Autoethnography}. \citet{gonzalez_CHI} investigate responses to the trustworthiness of AI descriptions in current tools. \citet{bennett_itscomplicated} explore the risks of AI bias when AI is used to auto-describe images, specifically photographs. Through an auto-ethnographic study, \citet{Glazko_Autoethnography} highlight how a BLV researcher wants to use AI for writing and validating code. While these research efforts provide a foundation for future AI-based visual artifact understanding systems, much of this prior research focuses on BLV adults' \textit{individual} usage of and interaction with AI.

Most relevant to our work, \citet{chhedakothary2024} performed a formative study exploring mixed visual-ability families' reactions to existing AI tools when used for interpreting children's artwork, finding generally positive reactions to AI descriptions. However, the researchers also highlight that families want to \textit{correct} inaccurate AI interpretations, as well as prevent reductive or overly simplistic AI language when describing their children's artwork. As formative work, \citet{chhedakothary2024} further inform our design guidelines for the ArtInsight system. 

Another category of recent research involves \textit{implementing} novel AI-based systems for BLV people to engage with visual content \cite{Huh_GenAssist, Kim_DescriptionsForComics}. \citet{Huh_GenAssist} enable accessible ways for BLV people to create AI-generated images, and \citet{Kim_DescriptionsForComics} compare AI and human annotations of comic strip descriptions. Both of these works leverage AI for the accessibility of different types of visual artifacts, but the focus of ArtInsight as a system for children's artwork interpretation remains unique. Additional research leverages prototypical systems using state-of-the-art AI models for physical world navigation tasks, such as helping BLV people find their personal belongings \cite{Morrison_FindMyThings} and with street crossings \cite{jain2024streetnavleveragingstreetcameras}. A technique that \citet{Morrison_FindMyThings} employ for training AI models to recognize BLV people's personal items is few-shot learning, which allows AI models to \textit{``make accurate predictions by training on a very small number of labeled examples.''} \footnote{https://www.ibm.com/topics/few-shot-learning} We similarly employ few-shot learning to evaluate the effectiveness of ArtInsight's prompt-engineered AI model across a small dataset of example children's artworks.

\subsection{Augmenting AI Descriptions with Context and Human Edits}
As \citet{chhedakothary2024} and \citet{bennett_itscomplicated} report, there is a desire by BLV individuals and their families to feed context and corrections to AI descriptions of visuals such as artwork or images of people. Based on this, we ground our work in prior research in context-aware AI \cite{Gubbi_ContextAware} as well as enabling human corrections or additions to AI descriptions \cite{be_my_eyes_2024, darth_vader}. We draw on efforts in the realm of BLV accessibility \cite{Lee_RSAs, be_my_eyes_2024, Gubbi_ContextAware, Singh_FigureA11y} as well as in the broader Human-AI research space \cite{darth_vader, hwang_creative}.

Even after its AI integration, \textit{Be My Eyes} continues to support human sighted assistance---Be My AI users can connect with human volunteers for further help. \citet{Gubbi_ContextAware} take a fully automated approach, creating a pipeline to extract meaningful and relevant data from websites to provide users with context-aware image descriptions. \citet{darth_vader} explore children's attitudes towards AI, and report findings that children want to provide added context to AI when it is incorrect or it misinterprets their request. We evaluate these different approaches as considerations for how ArtInsight should augment AI descriptions of artwork with the children's \textit{i.e.,} the artist's context and interpretation of their art.

\section{Design Considerations for Interpreting Child Artwork}
\label{DGs}
Drawing on relevant prior work, generative AI design principles \cite{GAI_DesignPrinciples}, and our own experiences building and testing initial prototypes, we synthesized a set of design considerations for AI-based child artwork interpretation to support mixed visual-ability families. We first enumerate design considerations around Human-AI systems, rooted in research on promoting dialog between family members and children \cite{Dietz_ContextQ, Zhang_StoryBuddy}, BLV people's preferences around understanding visual artifacts \cite{Li_UnderstandingVisualArtsExperiences, bennett_itscomplicated}, children's experiences with AI systems \cite{darth_vader}, and formative work on AI for artwork interpretation in mixed visual-ability families \cite{chhedakothary2024}. 

\begin{itemize}
    \item[DG1] The system should center the narrative and interpretation of the artwork creator \textit{i.e.} the child. As a part of this, the system should allow for child-added context or corrections \cite{darth_vader, Li_UnderstandingVisualArtsExperiences, chhedakothary2024, GAI_DesignPrinciples}.
    \item[DG2] The system should be \emph{supplementary} to the interaction between the family member and child---to this end, the system should help \textit{begin} (and potentially help sustain) family discussions \cite{Dietz_ContextQ, Zhang_StoryBuddy}.
    \item[DG3] The system should support different temporal stages of interaction with the child: before, during, and after co-exploring the artwork \cite{Zhang_StoryBuddy, chhedakothary2024}. 
    \item[DG4] The system should allow BLV family members to experience a more creative interpretation of child-created artwork that is still factual \cite{Li_UnderstandingVisualArtsExperiences, chhedakothary2024}.
\end{itemize}

Furthermore, informed by work investigating AI descriptions of visual artifacts for BLV people \cite{bennett_itscomplicated, Kim_DescriptionsForComics, Das_ProvenanceToAberrations, chhedakothary2024}, we additionally present four AI-specific design goals for generated descriptions of children's artwork.

\begin{itemize}
    \item[AI DG1] AI descriptions should not be presumptive---the AI agent should refrain from making overreaching assumptions about the artwork's content \cite{chhedakothary2024, bennett_itscomplicated}.
    \item[AI DG2] AI descriptions should not be reductive---the AI agent should avoid language that minimizes the effort or drawing style of the child \cite{chhedakothary2024}.
    \item[AI DG3] AI descriptions should offer detailed summaries, not simplistic descriptions, of children's artwork \cite{Das_ProvenanceToAberrations, Kim_DescriptionsForComics}.
    \item[AI DG4] AI descriptions should capture all major elements of the artwork, including holistic artwork characteristics and all visible elements \cite{chhedakothary2024}.
\end{itemize}

\section{ArtInsight Design and Implementation}

\begin{figure*}[h]
\centering
\includegraphics[width=\linewidth]{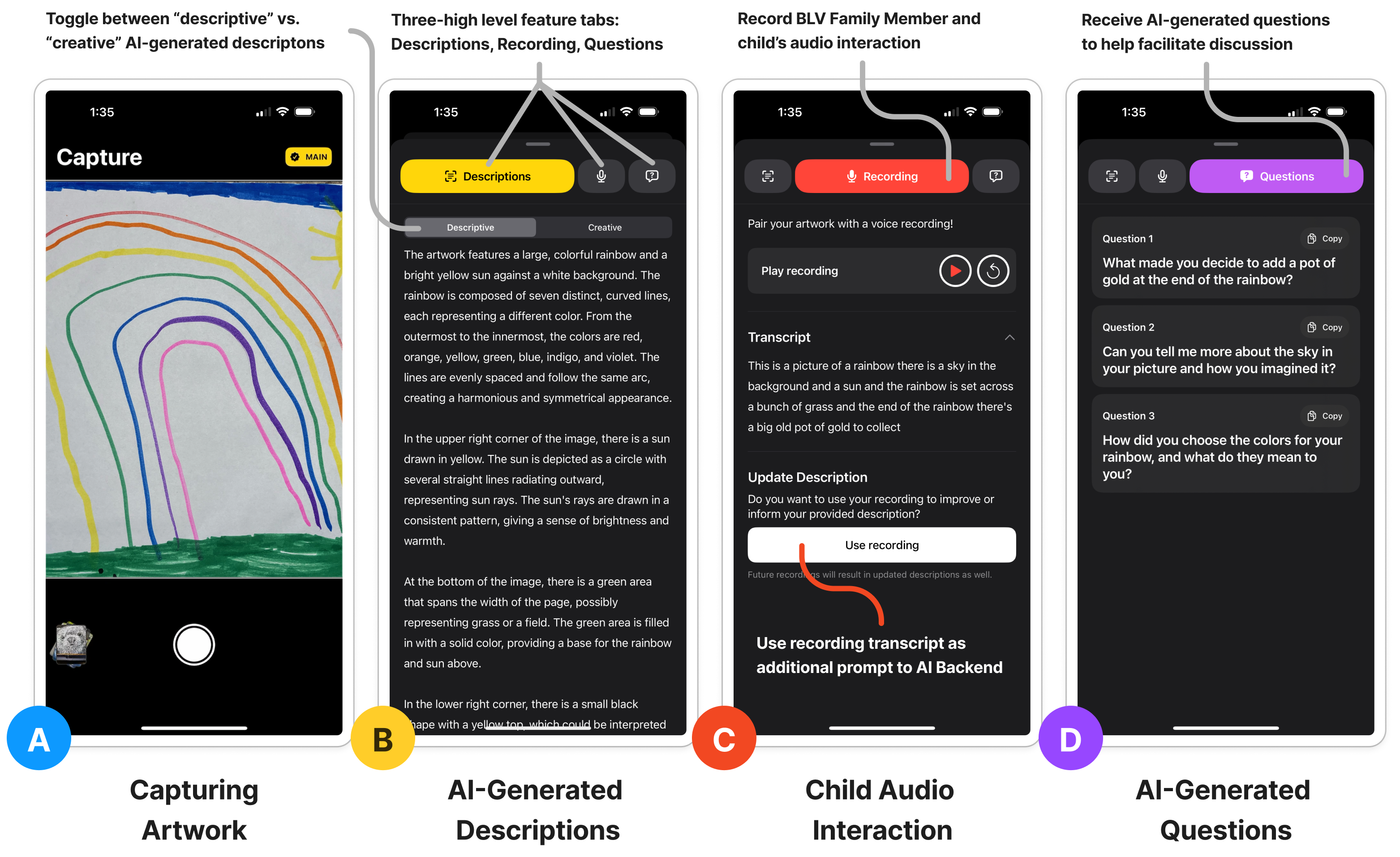}
\Description{A system design overview consisting of four screenshots of the ArtInsight application. The first pane on the far left is of the 'capture' stage, showing the mobile prototype taking a picture of a child's artwork with a rainbow. The second pane (center left) shows the "Descriptions" tab, which has some text and toggles for a descriptive and creative description. The third pane (center right) shows the "Audio" tab, which has a child's stored audio with buttons to play or retake the audio, and the transcript of the stored audio with a button to "Use recording" to re-generate AI descriptions. The fourth pane (far right) is of the "Questions" tab, with three AI-generated questions about the rainbow picture shown on the screen. Each screenshot has some corresponding labels describing the related functionality.}
\caption{ArtInsight allows BLV family members to (A) take photos of child-created artwork, (B) hear AI-generated descriptions of the artwork, (C) record and use an audio recording of the child describing their artwork for added context sent to the model, and (D) parse three AI-generated starter questions for dialogue and discussion with children.}
\label{System}
\vspace{-0.5em}
\end{figure*}

\begin{figure*}[h]
\centering
\includegraphics[width=\linewidth]{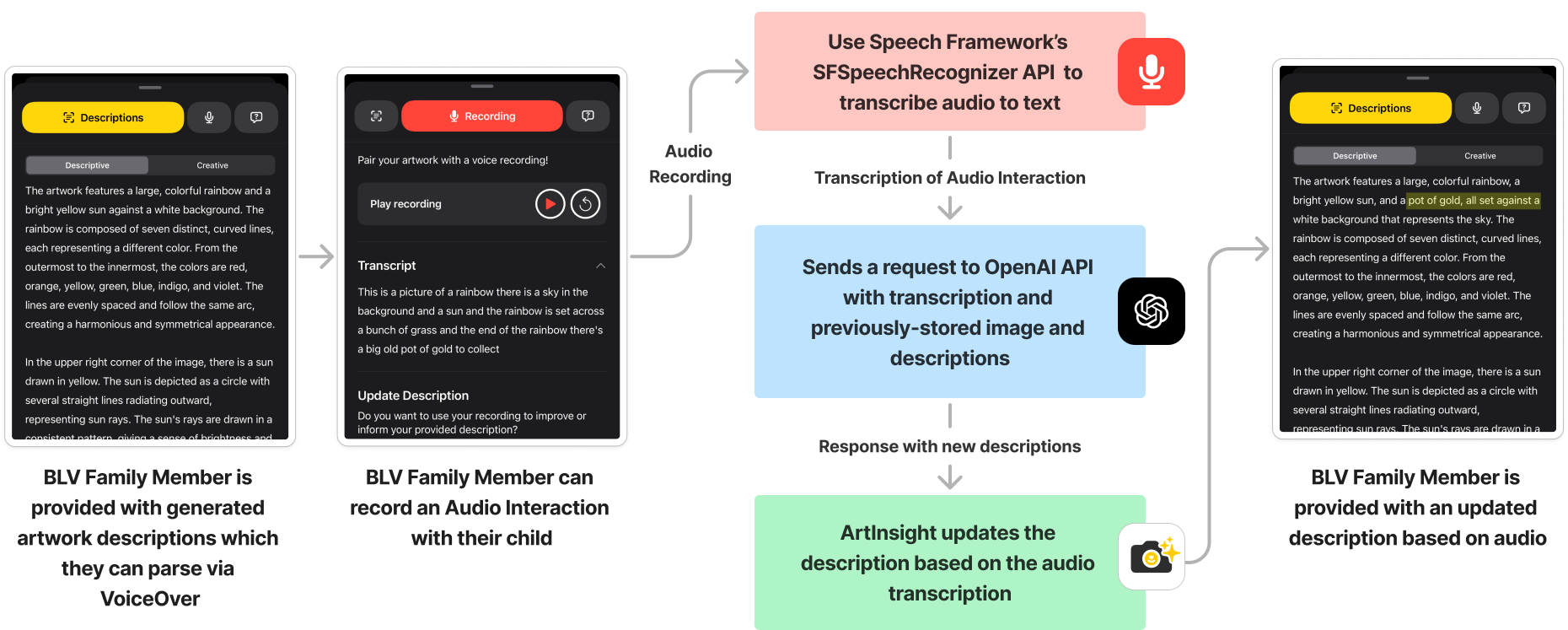}
\Description{The diagram shows through screenshot panels and a flow diagram how ArtInsight updates artwork descriptions for blind or low-vision (BLV) family members through audio interactions.
Left Panel: A BLV family member receives an initial AI-generated artwork description, parseable via VoiceOver.
Middle Panel: The family member records an audio interaction with their child describing the artwork, and the system transcribes it.
Right Panel: The updated description, based on the audio recording, is provided to the BLV family member.
In between the middle and right panel, there is a flow digram depicting the audio transcription process through the Speech Framework's SpeechRecognizer API, the transcript being sent to the OpenAI API, and ArtInsight receiving the new description from OpenAI based on the transcription. The flow diagram goes from top to bottom, with three boxes; first depicting the Speech Framework work (colored red), second depicting the OpenAI request (colored blue), and third depicting the ArtInsight updated description (colored green).}
\caption{The end-to-end workflow of navigating to the audio recording component, capturing an audio snippet, and using the transcript of the snippet to update the AI description of the artwork.}
\label{Audio}
\vspace{-0.5em}
\end{figure*}

Informed by the above design goals, we designed and built \textit{ArtInsight}, an AI-based prototype that uses large language models (LLMs) to enable understanding of child-created artwork and facilitate discussion in mixed visual-ability groups. To deepen interactions between the child and their BLV family member while centering human agency, ArtInsight allows BLV family members to record an optional audio note of the child's description of their work. This audio description is used to further customize the LLM description (Figure \ref{Audio}), to control the "personality" of the AI-generated output (\textit{e.g.,} "creative" \textit{vs.} "descriptive" descriptions), and to receive AI-generated questions in support of additional dialogue (drawing on prior work such as \textit{ContextQ} \cite{Dietz_ContextQ}).

ArtInsight is designed to support three flexible child-family member stages of access per DG3: a potential independent examination by the BLV family member preceding conversation with the child, \textit{in situ} interactions with the child and family member co-exploring the artwork together, and \textit{post hoc} investigations following the co-exploration (\textit{e.g.,} the family member re-examines the artwork independently). To use ArtInsight, BLV family members---possibly aided by sighted partners or children---take a photo of the child's artwork. Using a prompt-engineered version of GPT-4o \cite{explore_gpts_2024} as the backend AI agent, ArtInsight analyzes the image and provides an initial AI-generated description. This initial description (\textit{i.e.,} the "descriptive" description) is accessible both as visible text and as screen reader content. To facilitate additional interpretation and dialogue with children, BLV family members can toggle the "creative" description of the art, capture an audio recording of their child's description of the art and use this recording to update the AI description, and explore AI-generated questions about the artwork. ArtInsight locally stores all captured images, AI-generated descriptions, and audio recordings for future perusal.

The ArtInsight user experience was informed by existing BLV AI exploration apps such as Be My AI \cite{be_my_ai_2024} and Seeing AI \cite{seeing_ai}, as well as accessible design resources such as Apple's Human Interface Guidelines \cite{hig_accessibility}. We built ArtInsight with the SwiftUI \cite{swiftui} and UIKit \cite{uikit} frameworks for iOS, utilizing OpenAI's Assistants API \cite{openai_assistants} for the AI backend. To support mixed visual-ability families, we designed ArtInsight with full support for VoiceOver, iOS's built-in screen reader, \cite{voiceoverios} alongside the default touchscreen input controls. Below, we describe key ArtInsight components.

\textbf{Taking a picture.} To begin, users open the ArtInsight application that defaults to a photo-capture screen (Figure \ref{System}A). We based the photo-capture experience on popular BLV image description applications such as Be My AI and \textit{Seeing AI}. Guided by VoiceOver labels, BLV family members can either independently take the photo of their children's artwork based on the VoiceOver feedback about the capture button, or with sighted adult or child assistance.

\textbf{Initial description.} After the photo capture, ArtInsight sends a request to the OpenAI API \cite{openai_assistants} (the API layer for communicating with a custom prompt-engineered GPT-4o agent as our backend AI) with the image. The image is also simultaneously saved locally. The backend AI takes between 20-25 seconds to return a response containing two descriptions (a "descriptive" description and a "creative" description), an AI-generated title for easy perusal and retrieval of images stored locally, and AI-generated questions about the image. The AI-generated "descriptive" description is used as the initial description that the system displays as text, which can also be read aloud via VoiceOver (Figure \ref{System}B). We discuss the process to arrive at the custom AI prompt more thoroughly in Section \ref{AIBackend}.

\textbf{Controlling AI personality.} As described, for each photo, two AI descriptions are generated: "descriptive" and "creative" (Figure \ref{System}B). The creative description, informed by DG4, is generated by the same custom prompt powering the descriptive description, but with added instructions to allow the model to be more presumptive and vivid with its interpretation of the art. The user can toggle between the descriptive and creative descriptions on the Descriptions tab.

\textbf{Audio recording.} Though children are not always available when a BLV family member is experiencing their artwork, \citet{chhedakothary2024} found that the child's own perspective and story about the art is paramount. Thus, drawing on DG1---which emphasizes the child's narrative of their artwork---ArtInsight enables recording a snippet of audio so children can describe their artwork. The BLV family member can playback this recording directly (Figure \ref{System}C, the Audio tab). Moreover, to further customize the backend AI prompt, the audio recording is transcribed using iOS's Speech framework \cite{speech_framework} and fed into the model. Subsequently, the "descriptive" and "creative" descriptions along with the AI-generated questions get updated to reflect the new context (Figure \ref{Audio}). We chose audio as a low-friction, casual medium \cite{Chen_AudioText, MagicCamera} to capture children's descriptions of their art as well as any interactions between the parent and the child that families wish to record.

\textbf{AI-generated questions.} Finally, to help stimulate and strengthen discussion (DG2), ArtInsight provides an AI-generated set of example questions for BLV family members to ask their children. These questions (accessed from the Questions tab, Figure \ref{System}D) are generated through specific instructions sent to OpenAI as a part of the initial request when first analyzing the image, and get updated if AI descriptions are regenerated based on an audio recording.

\section{AI Model Selection and Evaluation}
\label{AIBackend}

While the above section described the key design and interaction components of the ArtInsight interface, we now describe the AI backend and model selection process powering ArtInsight. We designed an experiment to compare the quality of four state-of-the-art LLMs---Claude 3.5 Sonnet \cite{Claude35}, GPT-4 Turbo \cite{GPT-4Turbo}, GPT-4o \cite{GPT-4o}, and Gemini 1.5 Flash \cite{Gemini}---in generating AI descriptions of child artwork. To assess model output, we operationalized the AI design goals in Section \ref{DGs} into a domain-specific scoring rubric. To help scale our experiment, we used an automated scoring model, called \textit{LLM Scorer}. LLM Scorer is a GPT-4o \cite{GPT-4o} model, custom-trained using few-shot learning. We describe how we trained LLM Scorer and use it to compare the four state-of-the-art models. Finally, we describe our experimental findings, the best performing model, and how we arrived at our final prompt used for ArtInsight.

\begin{table*}[h]
  \centering
  \renewcommand{\arraystretch}{1.5}
  \normalsize
  \resizebox{\textwidth}{!}
  {
    \begin{tabular}{p{0.12\linewidth} p{0.34\linewidth} p{0.27\linewidth} p{0.27\linewidth}}
      \hline
      \textbf{AI Design Goal} 
      & \textbf{Rubric Guideline} 
      & \textbf{Example Low Score with} \hfill 
      
      \textbf{Rationale} 
      & \textbf{Example High Score with} \hfill
      
      \textbf{Rationale} \\
      \hline
        AI DG1: \hfill 
        
        Descriptions should not be presumptive.
        &
        A. Is the description being presumptive, \textit{i.e.} when it doesn't know something is it making inferences or assumptions about what they could be? For ex.: ``The main figure in the artwork is a large, dark gray shape in the center. It's hard to say for sure what it is, but it might be a person or animal.''---Ideally the description should say: ``the main figure in the artwork is a large, dark gray shape in the center.''
        & 
        1/4: The description makes several assumptions, such as suggesting the main figure might be a ``person or animal'' and that the word ``HOR'' might be part of a word. It also asks the parent what the child was thinking, which is speculative.
        &
        4/4: The description avoids making assumptions about the intent behind the artwork. It focuses on describing the visible elements without inferring any underlying messages or reasons.
        \\
        \hline
        AI DG2: \hfill 
        
        Descriptions should not be reductive.
        &
        B. Is the description being reductive, \textit{i.e.} is it ever minimizing the effort or drawing style of the child? For ex., a description that says, ``this is a drawing of simple stick figures'' can cause the parent to dislike the use of the word ``simple.'' Another example: ``this is a rough rectangle''---descriptions that use terms like `rough' diminish the work the child has put in.
        & 
        2/4: The description uses phrases like ``it's hard to say for sure,'' which can come across as dismissive. It does not fully appreciate the effort and creativity of the child.
        &
        4/4: The description is respectful and acknowledges the emotional depth of the artwork, using terms like ``interesting and powerful piece'' and ``shows a lot of feeling.'' It does not use any diminishing language.
        \\
        \hline
        AI DG3: \hfill
        
        Descriptions should offer detailed \hfill
        
        summaries.
        &
        C. Is the description too simple, i.e. only saying things like: ``This is a child's drawing of a forest and some animals.'' Ideally the description goes into detail about the artwork.
        & 
        2/4: The description is somewhat simple and lacks depth. It mentions the main elements but does not go into detail about the texture or the overall feel of the art.
        &
        4/4: The description is detailed and covers several aspects of the artwork, such as colors, shapes, textures, and the combination of drawing and crafting elements.
        \\
        \hline
        AI DG4: \hfill
        
        Descriptions should capture all major elements of the artwork.
        &
        D. Are all the major elements of the artwork captured?
        & 
        2/4: The description captures some major elements but misses the detail about the specific letters ``HBD'' in the artwork. It also does not mention the yellow color in the palm area of the handprint.
        &
        4/4: The description captures all the major elements: the rectangular shape, the pom-poms, the green pipe cleaner, the heart shapes, and the text at the bottom right. It provides a comprehensive view of the artwork’s details.
        \\
      \hline
    \end{tabular}
  }
  \caption{The AI Design Goals, corresponding Rubric Guidelines, and examples of low and high scores with corresponding explanations of the scoring. All rubric categories are scored on a 0-4 scale, and there is an additional miscellaneous subtraction allowance for any description language or errors detracting from the quality of the description.}
  \label{Rubric}
  \vspace{-1.25em}
\end{table*}

\textbf{Rubric and initial prompt.}
Before we could conduct a thorough comparison of state-of-the-art multimodal LLMs for describing child-created visual artwork, we needed to establish a guide for determining an effective description. Though there are general benchmarks for image descriptions \cite{GPT-4_2023, Gupta2022GRITGR}, there are none specific to evaluating children's artwork. Furthermore, AI descriptions of children's artwork should adhere to certain best practices---they should be respectful, thorough, and avoid interpretation (unless the artist's interpretation is explicitly given) \cite{chhedakothary2024}. 

Based on this, we operationalized our AI design goals into a formal rubric (Table \ref{Rubric}), following rubric-based evaluation of AI practices from prior research \cite{RUBICON}. We use a variation of a 5-point Likert scale for scoring, adjusted to a 0-4 scale instead of 1-5 to allow for a minimum score of 0. Beyond the guidelines listed in Table \ref{Rubric}, our rubric allows for a "Miscellaneous" subtraction of points: \textit{``Are there any other parts of the response which take away from the overall quality?''} Based on our rubric, we crafted the following initial prompt for the model comparisons:

\begin{quote}
    \textit{This assistant's name is Art Insight. Art Insight helps blind parents understand their children's visual artwork. It provides detailed, respectful descriptions of the artwork, focusing on descriptive aspects such as orientation, scenery, number of artifacts or figures, main colors, and themes.} \\
    \textit{The assistant avoids reductive or overly simplifying language that minimizes the child's effort and does not assume interpretations if uncertain. For example, it says, `The person has a frown, and there are tears falling from their eyes' instead of `The person appears to be sad.' When given feedback from the parent or child about the artwork, the assistant honors and integrates this perspective into its descriptions and future responses.} \\
    \textit{The assistant maintains a respectful, supportive, and engaging tone, encouraging open dialogue about the artwork. Art Insight uses a casual tone but will switch to a more formal tone if requested by the parent or child. The assistant avoids making assumptions about names or identities based on any text in the artwork. The response should be in paragraph form.}
\end{quote}

\textbf{LLM Scorer.}
While the rubric enabled us to score model responses, we wanted to additionally automate the scoring process to be able to score responses \textit{at-scale} for our model comparison experiment. For this, we created the \textit{LLM Scorer}---a custom GPT-4o \cite{GPT-4o} model, trained using few-shot learning \cite{parnami2022learning, Morrison_FindMyThings}. We employed few-shot learning as a method because of the limited number of children's artworks we had access to to use as training data. We chose GPT-4o as the model for the LLM Scorer as it was the leading vision-language model at the time of our experiment \cite{(HELM)}.

To train the LLM Scorer, the lead researcher first collected five diverse sample artwork images (brushstroke paintings, notebook doodles, crayon coloring projects) created by children ages 4-17 from friends and family. We ran those five images through the four state-of-the-art models chosen for our experiment with our initial prompt. Two researchers from our team manually scored each of the four models' image descriptions using our rubric, and documented rationale for their scores such as: \textit{``4/4: The description is detailed and covers aspects of the artwork including colors, shapes, brushstrokes, and the composition.''} These five images along with their corresponding descriptions, scores, and documented rationale were then used to prepare the LLM Scorer. The input to the LLM Scorer is an image and an AI-generated description, and the output from the LLM Scorer is a final score (out of 16 points, per our rubric) and the scoring rationale across the different rubric categories.

\textbf{Model comparison.}
For our experiment, we compared Claude 3.5 Sonnet, GPT-4 Turbo, GPT-4o, and Gemini 1.5 Flash. We chose these models because at the time of the experiment, they were among the highest-performing multimodal LLMs on tests and benchmarks such as Massive Multitask Language Understanding and Image2Struct \cite{(HELM), HELM_MMLU}, which provided good measures for general image and task understanding. The lead researcher first curated a dataset of 30 images of artwork created by children ages 3-17 from friends, family, and BLV community members, who provided consent to use their children's art for our work. We then used our initial prompt to generate descriptions for all 30 images across the four models, producing a total of 120 image descriptions.

To evaluate these 120 descriptions, we employed the LLM Scorer. Members of our research team spot-checked the scores and explanations produced by the LLM Scorer to correct for any hallucinations or errors---four members of the research team each checked five unique description scores (randomly chosen and assigned).

\textbf{Results.}
The GPT-4o and Claude 3.5 Sonnet models performed best with average scores of 15.7/16 and 14.9, respectively, followed by GPT-4 Turbo (14.7) and Gemini 1.5 Flash (11.5). To help illustrate performance differences, Appendix Table \ref{MasterTable} provides three example children-created artworks along with their final scores and the explanation of the points subtracted by the LLM Scorer and human spot-checkers. We also provide the full set of scores in Appendix Table \ref{Test}. Based on these findings, our backend used GPT-4o.

\textbf{Final prompt.}
Finally, we used the LLM Scorer to help evaluate changes to our initial prompt, \textit{i.e.,} additional prompt engineering, using the 30-image dataset from the model comparison experiment. Our final prompt (Appendix \ref{FinalPrompt}) with GPT-4o as the artwork description model produced an average score of 16.0 from the LLM Scorer across the 30 images from the dataset. Our research team again spot-checked the descriptions generated by GPT-4o with all iterations of our prompts (including our final prompt) to correct for LLM Scorer errors and hallucinations.

\section{User Study with Mixed Visual-Ability Families}
To evaluate ArtInsight and gain insight from BLV adults and their sighted children, we conducted an in-person user study with five groups of mixed visual-ability families.

\subsection{Participants}
We recruited five mixed visual-ability families with sighted children via email lists, social media, and snowball sampling \cite{snowball_sampling}. The BLV adult participants consisted of four parents and one grandparent. See Table \ref{Participants} for demographics.

\subsection{Apparatus}
For each study session, we used an iPhone 12 Pro (iOS 17.6.1) with both ArtInsight and Be My AI \cite{be_my_ai_2024} (the AI mode of Be My Eyes, version 5.3.18) installed and VoiceOver enabled. Participants were invited to bring two or more of their own children's 2D visual artworks to use during the session. We placed no restrictions on the artwork except that it must be 2D (\textit{i.e.,} no 3D constructions like clay models or tactile-heavy artwork). We did not specifically ask participants to bring artwork that had not been previously discussed with their children.

\subsection{Procedure}
Our study consisted of three parts: (1) a comparison between the initial AI descriptions from ArtInsight and Be My AI, (2) an exploration of key ArtInsight components (\textit{e.g., audio recording}), and (3) a semi-structured interview. Sessions lasted for ~90 minutes. Participants were compensated either \$25/hr. if the session occurred at a participant's house or \$40/hr. if a participant traveled to the study location. Prior to the study session, participants completed a pre-study questionnaire, which asked background information about their family and vision loss. Participants were also asked to bring in artwork created by their children. After arriving, participants were provided with the consent statement, an overview of the study, and a tutorial of ArtInsight and Be My AI using an example artwork provided by the lead researcher. Both the BLV family member and their child(ren) participated throughout the entire study session. We expand on the three parts of the study, below.

\textbf{Part 1: ArtInsight \textit{vs.} Be My AI}. First, participants heard descriptions of their child's artwork from ArtInsight and Be My AI using VoiceOver \cite{voiceoverios}. We counterbalanced the use of each system across participants to prevent order effects. The BLV family members rated perceived accuracy and usefulness of descriptions on a 5-point Likert scale. We also asked the children semi-structured interview questions to understand their responses to the different descriptions after the BLV family members rated the descriptions.

\textbf{Part 2: Exploration of ArtInsight features}. Second, participants examined the three novel ArtInsight components: the creative interpretation toggle, audio recording with AI re-prompting, and the AI-generated questions. We specifically asked participants to switch the descriptive and creative interpretation toggle, to record an audio description of the child explaining their artwork, and to navigate to the questions tab to parse the three AI-generated questions. After using each component, the family groups answered questions about their experiences. BLV family members and their children chose \textit{one} of the artworks they brought for testing the three components.

\textbf{Part 3: Semi-structured interview}. Third, we conducted semi-structured interviews to solicit holistic feedback on ArtInsight. We also asked participants to rank the three novel ArtInsight components from most to least preferred.

\subsection{Analysis}
We recorded and transcribed all study sessions. To analyze the comparison portion of the studies for which we had Likert scale ratings, we used mixed ordinal logistic regression \cite{GLMM, OrdinalRegression}. We also engaged our LLM Scorer and three human scorers (two from our team, one from an external AI research group) to use our rubric to evaluate the initial descriptions of participant-provided artwork created by ArtInsight and Be My AI. For qualitative analysis, two researchers from our team used deductive and inductive coding \cite{Braune_And_Clarke} with reflexive thematic analysis \cite{ThematicAnalysisHCI} to arrive at an initial set of 13 codes. The lead researcher then conducted peer debriefing sessions \cite{PeerDebriefing} with the two researchers to refine the codes.

\begin{table*}[h]
\resizebox{0.95\linewidth}{!}{ 
\begin{tabular}{ccclcl}
\toprule 
\rowcolor{white}  
\textbf{PID} & \textbf{Age Range} & \textbf{Gender} & \textbf{Degree of Vision Loss} & \textbf{Age of Children} & \textbf{Relationship to Children} \\
\midrule

1 & 45-54 & F & Totally Blind/No Usable Vision & 17 & Mother \\

2 & 45-54 & M & Legally Blind & 7 and 13 & Father \\

3 & 25-34 & F & Legally Blind & 6 & Mother \\

4 & 25-34 & M & Legally Blind & 13 & Father \\

5 & 55-64 & F & Totally Blind/No Usable Vision & 9 & Grandmother \\

\bottomrule
\end{tabular}
}
\vspace{0.5em}
\caption{Self-reported demographics of BLV family members and children who participated in the first study.}
\label{Participants}
\vspace{-1em}
\end{table*}

\section{Study Findings}
Our user study revealed that participants preferred the detail and structure of ArtInsight’s initial AI responses over those of Be My AI, with additional positive feedback highlighting the creative interpretation and audio recording components as helpful to feel closer to \textit{``what the artist was thinking''} (P1). Below, we compare ArtInsight descriptions to Be My AI descriptions, discuss reactions to each ArtInsight component, and reflect upon how families envision using ArtInsight, their expectations of AI, and any suggested additions to ArtInsight.

\subsection{Part 1: ArtInsight \textit{vs.} Be My AI}
In Part 1, participants heard AI-generated descriptions of two pieces of artwork they supplied from both ArtInsight and Be My AI. We found that participants rated the \textit{Usefulness} of ArtInsight descriptions higher than those of Be My AI, but the \textit{Perceived Accuracy} of descriptions across the applications was similar.

\begin{figure*}[h]
\centering
\includegraphics[width=0.95\linewidth]{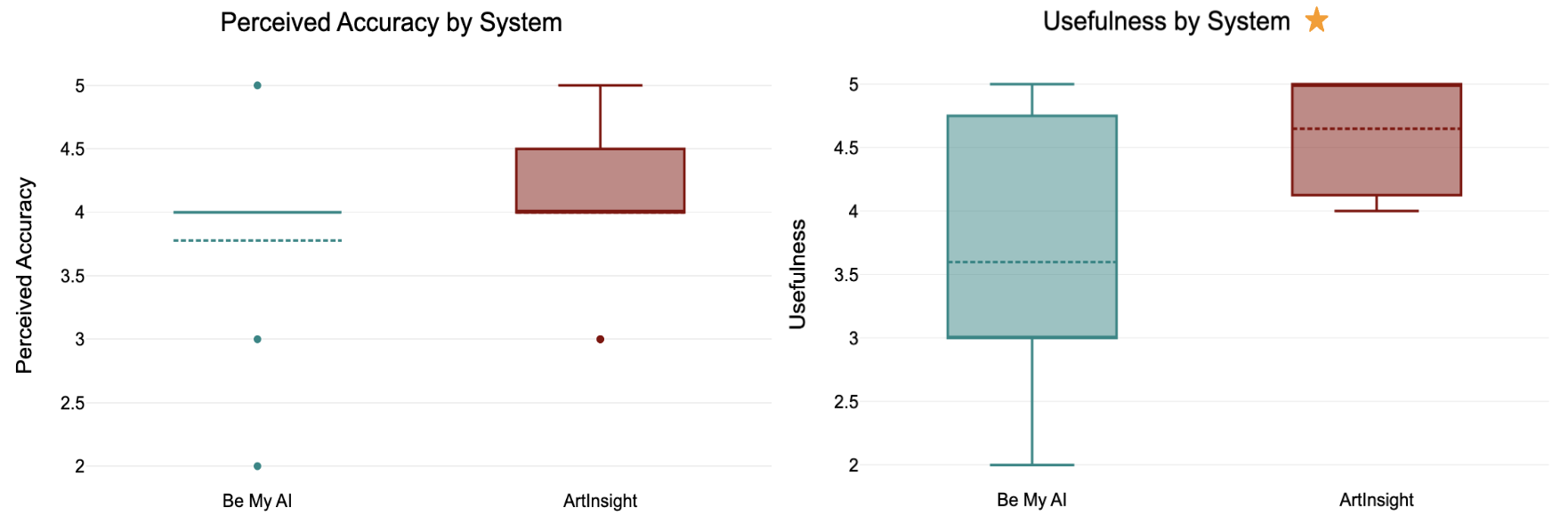}
\Description{Two box plots, one of perceived accuracy (left) and one of Usefulness (right). Usefulness has a star next to it indicating statistical significance. The box plots show visually that Perceived Accuracy across both systems was similar, while Usefulness for ArtInsight had much higher ratings than Be My AI.}
\caption{Participants' \textit{Perceived Accuracy} (not statistically significant) and \textit{Usefulness} (statistically significant) ratings for initial AI descriptions of Be My AI.}
\label{Usefulness}
\vspace{-1em}
\end{figure*}

\textbf{Usefulness}. Participants rated the \textit{Usefulness} of ArtInsight descriptions higher than Be My AI: $M$=4.7 ($SD$=0.5) \textit{vs.} $M$=3.6 ($SD$=1.1). This result was statistically significant ($\chi^2$(1, $N$=5) = 3.9, $p=.047$). When asked about the usefulness of ArtInsight \textit{vs.} Be My AI, participants highlighted the level of detail, the artistic language, and the structure of the descriptions from ArtInsight. For example, P1, after hearing both ArtInsight's and Be My AI's descriptions of her daughter's artwork of a ferret, said the ArtInsight description \textit{``was a lot more helpful and detailed...I feel like I can almost see it now.''} P2 said having more thorough descriptions of his children's artwork would enable him to \textit{``go and ask the kids more questions based off the longer description.''} P5 commented on being pleasantly surprised by the level of detail the ArtInsight descriptions provided, saying: \textit{``that's more descriptive than anything I have gotten on any [AI system] anywhere at any time.''} P4 appreciated the structure of ArtInsight's descriptions more than Be My AI's descriptions: 

\begin{quote}
    \textit{I liked that [ArtInsight] broke it down into segments. And then it came back around and summarized it. Like when you read a book, and then you summarize the whole entire chapter.} --- \textbf{P4}
\end{quote}

\begin{figure}[h]
\centering
\includegraphics[width=\linewidth]{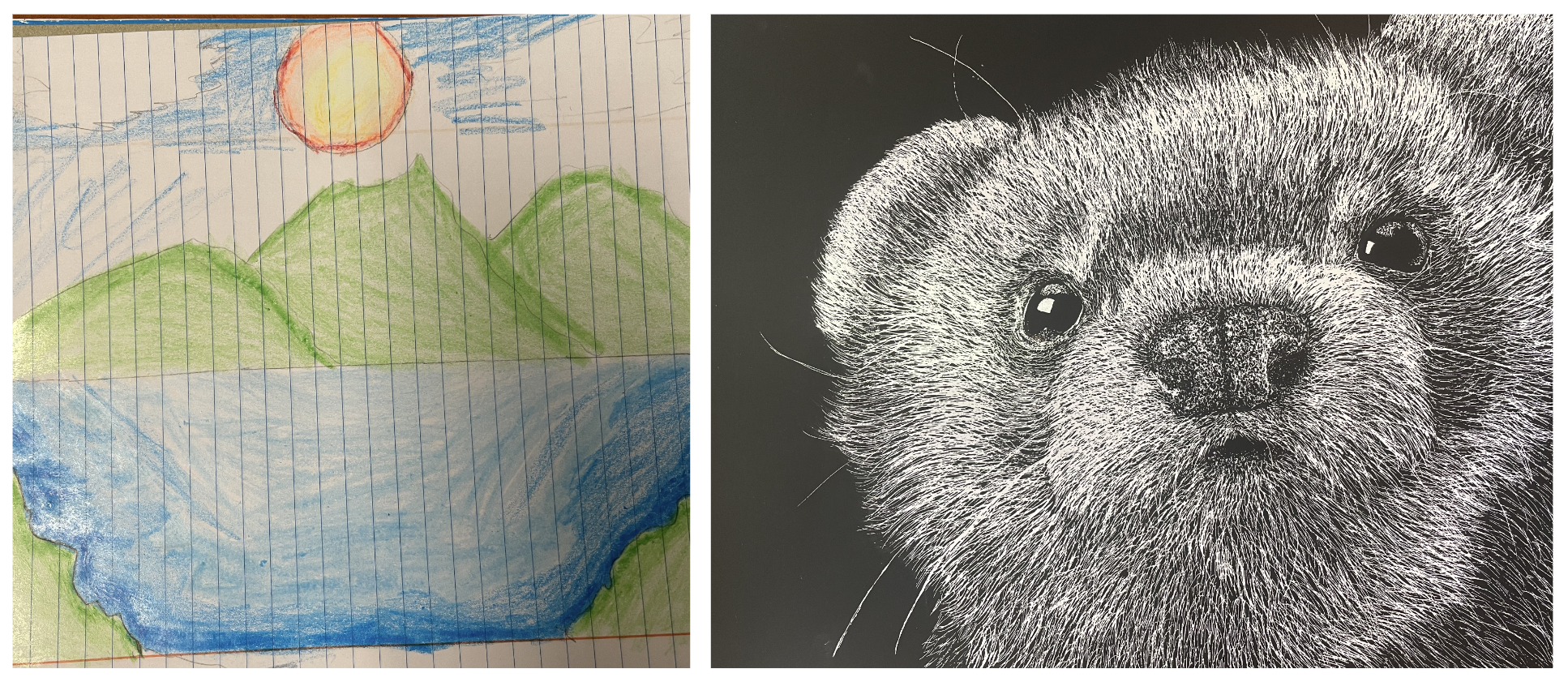}
\Description{Two child-created artworks; on the left, a sketch using colored pencils on a notebook paper of a mountain landscape with a blue lake, and a sun on the top. The shading work in the various landscape elements is detailed and careful. On the right, a monochromatic artwork of a ferret's face with extreme detail such that every stroke representing fur is discernible. The ferret looks surprised or intrigued.}
\caption{Example artwork by P4's son (\textit{left}) and P1's daughter (\textit{right}) used during the comparison portion of the studies.  }
\label{ShadingFerret}
\vspace{-1em}
\end{figure}

Children throughout the sessions similarly preferred the longer, more thorough descriptions provided by ArtInsight, and enjoyed the artistic language used by ArtInsight compared to Be My AI. P1's daughter liked the way ArtInsight described \textit{``the level of realism.''} P4's son appreciated ArtInsight's detailed description of the shading employed in his drawing (Figure \ref{ShadingFerret}, left), as he spent most of his time on the shading; on the other hand, Be My AI did not mention the shading at all.

\begin{figure*}[h]
\centering
\includegraphics[width=\linewidth]{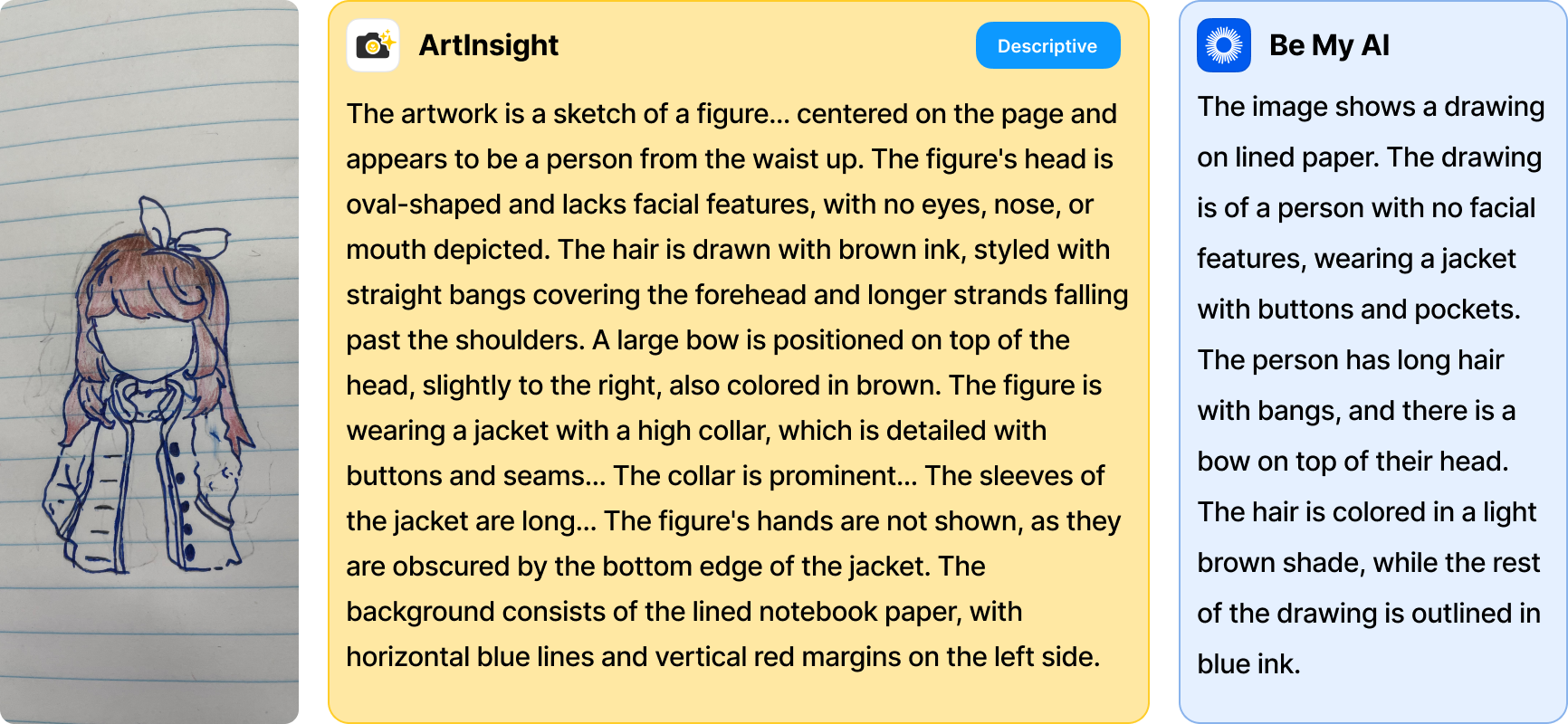}
\Description{A screenshot of three panels: on the left, a child's artwork of a person with no face details but with brown hair and a bow, a jacket outlined in blue, and no bottom half. In the middle: A snippet of the ArtInsight "descriptive" description, talking about the figure's head, hair, bow, jacket, high collar, jacket sleeves, and lined notebook paper background. On the right: The Be My AI description, saying: "The image shows a drawing on lined paper. The drawing is of a person with no facial features, wearing a jacket with buttons and pockets. The person has long hair with bangs, and there is a bow on top of their head. The hair is colored in a light brown shade, while the rest of the drawing is outlined in blue ink."}
\caption{P2's daughter's incomplete drawing of a person, along with snippets of the initial ArtInsight and Be My AI descriptions. Neither description mentioned the gradient of the hair, which P2's daughter wanted to surface for her father.}
\label{P2Daughter}
\vspace{-0.5em}
\end{figure*}

Many children felt that both ArtInsight and Be My AI missed important context about their artwork. For example, P2's daughter wanted her father to know that her drawing of a person was incomplete, something that neither description recognized. Furthermore, she had spent time on the gradient of the person's hair, but neither description mentioned the gradient (Figure \ref{P2Daughter}). In another example, neither system correctly identified the type of animal in P1's daughter's artwork (Figure \ref{ShadingFerret}, right). P1's daughter mentioned that if she described the artwork to her blind mother, she would always start by identifying the type of animal (a ferret).

\begin{figure*}[h]
\centering
\includegraphics[width=\linewidth]{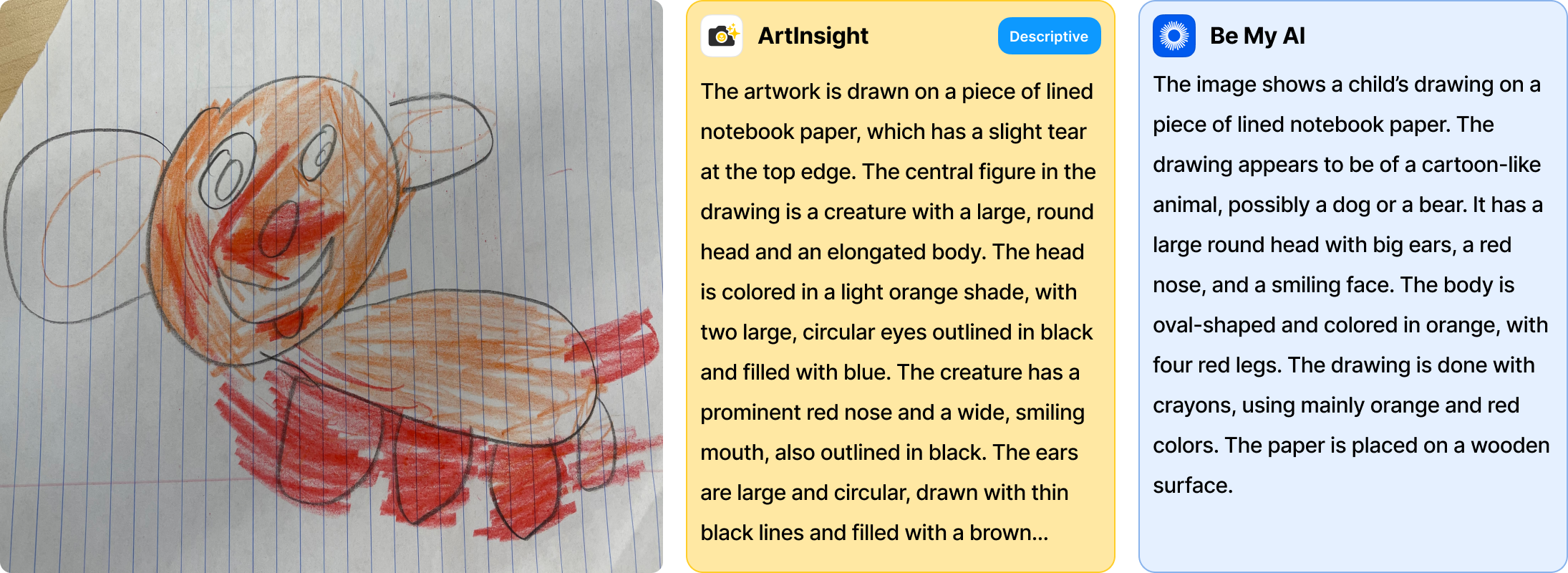}
\Description{A screenshot of P2's son's drawing of a dog, along with snippets of the initial ArtInsight and Be My AI descriptions. On the left: the colored pencil drawing of a dog with large round ears, a happy smiling face, and a body that is orange with red legs and a red tail. In the center: A snippet of the ArtInsight "descriptive" description, saying: "The artwork is drawn on a piece of lined notebook paper, which has a slight tear at the top edge. The central figure in the drawing is a creature with a large, round head and an elongated body. The head is colored in a light orange shade, with two large, circular eyes outlined in black and filled with blue. The creature has a prominent red nose and a wide, smiling mouth, also outlined in black. The ears are large and circular, drawn with thin black lines and filled with a brown..." On the right: The Be My AI description, saying: "The image shows a child's drawing on a piece of lined notebook paper. The drawing appears to be of a cartoon-like animal, possibly a dog or a bear. It has a large round head with big ears, a red nose, and a smiling face. The body is oval-shaped and colored in orange, with four red legs. The drawing is done with crayons, using mainly orange and red colors."}
\caption{P2's son's drawing of a dog, along with snippets of the initial ArtInsight and Be My AI descriptions. Be My AI attempted to guess the type of animal, which P2 and his son appreciated, compared to ArtInsight which called the dog a \textit{``creature.''}}
\label{DogScribble}
\vspace{-1em}
\end{figure*}

\textbf{Perceived Accuracy.} Ratings for \textit{Perceived Accuracy} did not show statistical significance ($\chi^2$(1, $N$=5) = 1.0, $p=.327$), indicating participants considered ArtInsight descriptions ($M$=4.0, $SD$=0.7) and Be My AI descriptions ($M$=3.8, $SD$=0.8)  similarly accurate. Anecdotally, some participants shared their preference for Be My AI descriptions that attempted to guess an unknown art element, even if the guess proved wrong. For example, ArtInsight described P2's son's drawing of a dog as a \textit{``creature''} (Figure \ref{DogScribble}). On the other hand, Be My AI's description said: \textit{``possibly a dog or a bear.''} P2 said that \textit{``even [if there was]... like a 50 percent chance... this might be a dog,''} he would want the AI to guess \textit{``dog.''}

Interestingly, P3 ranked Be My AI \textit{lower} (4) than ArtInsight (5) on accuracy \textit{because} of an assumption that Be My AI made, despite the fact that she liked the assumption: 

\begin{quote}
    \textit{I would give it a 4 because it did say that there was brown on the bottom representing the soil... I like that it made that assumption [but]... when it says representing soil, I imagine squiggly lines or something.} --- \textbf{P3}
\end{quote}

\textbf{Rubric-based performance evaluation.} As an additional comparison between ArtInsight and Be My AI, we used our rubric for quality AI descriptions of children's artwork (Table \ref{Rubric}) to score AI descriptions from the two applications during Part 1 of our Study across the \textit{LLM Scorer} and three human scorers. ArtInsight's initial description scores were generally higher than those for Be My AI---the mean LLM Scorer score for ArtInsight was 15.9 compared to 12.0 for Be My AI, and the mean score across the three human scorers for ArtInsight was 14.2 compared to 10.7 for Be My AI. The full set of scores from each scoring entity for ArtInsight and Be My AI are in Appendix Table \ref{Scoring}.




\subsection{Part 2: ArtInsight Components}
We share participants' reactions and feedback to the novel ArtInsight components---the creative interpretation toggle to access a more vivid artwork description, the audio recording to capture children's own descriptions and interpretations of their work with the ability to re-prompt the AI backend for a new description with context from the audio, and the AI-generated questions to start deeper artwork dialogue in families.

\textbf{Creative interpretation toggle.} Two participants, P1 and P2, ranked the creative toggle as their top of the three components; the other three listed the creative toggle as second. Participants appreciated the imagery that the creative descriptions evoked. P3 said the language was \textit{``a great choice... because it's kid's work, it's fun to hear [the creative description], it brings a picture to life.''} P1 and her daughter both thought the creative description felt more like \textit{``what the artist was thinking.''} P2 mentioned that he would want to start with the initial "descriptive" description of his children's work first, but he really enjoyed the creative description as it \textit{``gave [him] more of an idea of what it looks like.''}

\begin{figure*}[h]
\centering
\includegraphics[width=\linewidth]{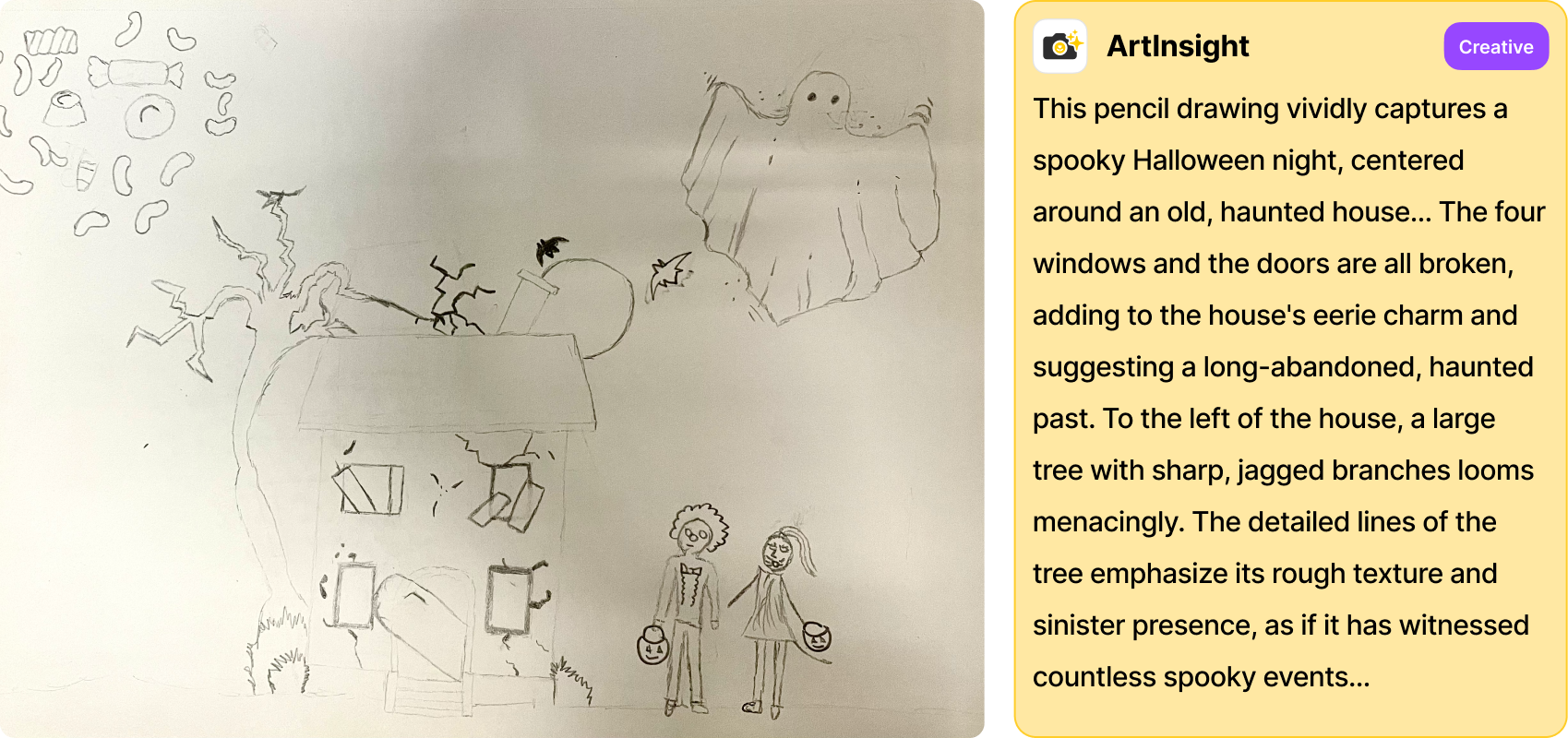}
\Description{On the left, there is a photo of a halloween-themed sketch. It contains a haunted house, a spiky tree, a moon and a floating ghost, candy floating in the top left corner, and a boy and girl trick-or-treating. On the right, there is a snippet of the ArtInsight "creative" description, which says: "This pencil drawing vividly captures a spooky Halloween night, centered around an old, haunted house... The four windows and the doors are all broken, adding to the house's eerie charm and suggesting a long-abandoned, haunted past. To the left of the house, a large tree with sharp, jagged branches looms menacingly. The detailed lines of the tree emphasize its rough texture and sinister presence, as if it has witnessed countless spooky events..."}
\caption{P4's son's Halloween drawing with a snippet of the AI-generated creative description. The creative description has language such as ``eerie charm'' and ``a long-abandoned, haunted past,'' creating a more striking rendition of the artwork.}
\label{Halloween}
\vspace{-0.5em}
\end{figure*}

We observed some discrepancies between P4 and his son when discussing the creative description. P4's son really enjoyed hearing the creative description of a Halloween-themed drawing he had done (Figure \ref{Halloween}), preferring it to the initial "descriptive" description because \textit{``it described how spooky and thrilling [the artwork] is... it felt like it was telling a story of the picture.''} However, P4 responded to his son's statement saying he thought the creative description was \textit{``too much flavor''} and \textit{``overkill.''}

\begin{figure*}[h]
\centering
\includegraphics[width=\linewidth]{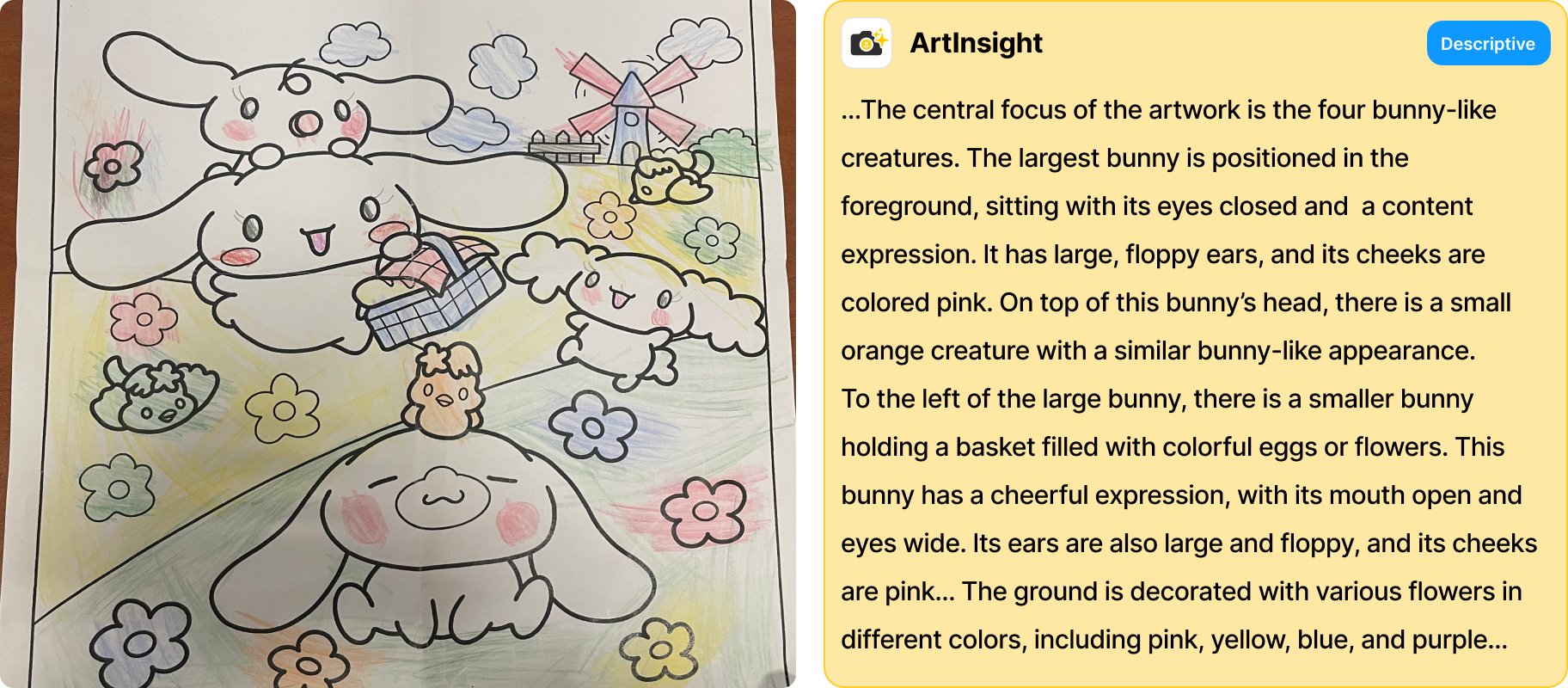}
\Description{On the left, there is a photo of an artwork done in the page of a coloring book with four bunnies who look like they are on a picnic, some birds scattered through the scene, a windmill, and many colorful flowers set on a pair of green and yellow hills. On the right, there is an ArtInsight "descriptive" description snippet associated with the artwork that says: "...The central focus of the artwork is the four bunny-like creatures. The largest bunny is positioned in the foreground, sitting with its eyes closed and a content expression. It has large, floppy ears, and its cheeks are colored pink. On top of this bunny’s head, there is a small orange creature with a similar bunny-like appearance.
To the left of the large bunny, there is a smaller bunny holding a basket filled with colorful eggs or flowers. This bunny has a cheerful expression, with its mouth open and eyes wide. Its ears are also large and floppy, and its cheeks are pink... The ground is decorated with various flowers in different colors, including pink, yellow, blue, and purple..."}
\caption{P3's daughter's coloring with a snippet of ArtInsight's ``descriptive'' description. P3's daughter immediately picked up on many of the small inaccurate details, such as \textit{``flowers''} in the basket instead of bread, or the (non-existent) color \textit{``purple.''}}
\label{P3Daughter}
\vspace{-1em}
\end{figure*}

\textbf{Audio recording with AI re-prompting.} Three participants (P3, P4, P5) ranked the audio recording with AI re-prompting component as their top, while the remaining listed it as their second favorite (P1) and least favorite (P2). P2 specifically did not find value in storing and using the audio recordings for image descriptions, preferring live interactions with his children to discuss their artwork. While he found the updated AI description based on the audio \textit{``cool,''} he personally wanted to pursue \textit{``other ways to capture what the kid meant''} outside of an AI tool. He also wanted to use ArtInsight exclusively \textit{before} interacting with his children, not during, which would hinder his ability to record the audio description.

P1 liked the audio recording, but considered it \textit{``redundant''} to have the AI description update for herself if she could access the raw audio of her child's description. However, she did want to share the updated AI description with her daughter's audio context:

\begin{quote}
    \textit{I could share the updated description... as alternative text on a photo that I'm posting, `Look at my daughters art,' and then my blind friends can read it.} --- \textbf{P1}
\end{quote}

For the participants that ranked the audio recording and AI re-prompting component as their favorite, they found value in being able to real-time correct and augment the original AI descriptions with their children's own descriptions. For these participants, being able to use their children's descriptions of their art in this manner also made ArtInsight as a tool feel more \textit{factual}. As P5 described: 

\begin{quote}
    \textit{[The audio recording component] added what [my granddaughter] added to the description... it gave [ArtInsight] more information that it needed to be accurate... [I would use this] to clarify detail and to get her opinion. What was her intention behind the picture? Where was she coming from?} --- \textbf{P5}
\end{quote}

Children also enjoyed the revised AI descriptions generated after they recorded their interpretation of their art. P5's granddaughter expressed that the AI description based on her audio recording \textit{``was a lot better''} than the original AI description. P4's son commented that he \textit{``liked how it [the description] changed,''} indicating an interest in the dynamic nature of ArtInsight. P1's daughter said the updated AI description based on her audio recording was \textit{``the best one,''} even saying: \textit{``I think [the description] better worded it than I did.''}

\textbf{AI-generated questions.} Four of the five participants (P1, P3, P4, P5) listed the AI-generated questions component at the bottom of their component rankings. The primary feedback was that as BLV parents and grandparents, participants typically already had questions of their own they would want to ask their children. As P4 put it: \textit{ ``I already generate my own 100 compliments and questions.''} P1 mentioned she found \textit{some} of the questions useful as they \textit{``inspired me to think of artistically framed questions,''} as her daughter is older and quite advanced at her craft. P3, in response to hearing all of the questions, said: \textit{``I don't really need this because I would have already asked [my daughter] these questions myself.''}

\subsection{Part 3: Debrief Interview}
We present results from our closing semi-structured interviews with families, including preferences for using ArtInsight and differences in expectations around AI between BLV adults and sighted children.

\textbf{When and how families want to use ArtInsight.} Participants expressed wanting to use ArtInsight primarily \textit{before} or \textit{during} their interaction with their child. P2 said he would `\textit{`probably use [ArtInsight] before [the interaction with the child] to get the idea of what the drawing is''} after which he would ask his children questions to learn more. He specifically did not want to use technology in the middle of interactions with his younger children, because he worried that it would distract them. P4 had a different perspective---he described wanting to use ArtInsight \textit{``on the spot''} as he discusses artwork with his son.

P1 wanted to actively use ArtInsight \textit{outside} of a dialogue with her child, for two main reasons---to avoid burdening her daughter for more details, and to share a thorough description of her daughter's artwork along with the artwork itself with her blind friends. Towards the first, P1 elaborated:

\begin{quote}
    \textit{[My daughter can say] `Mom, I'm sending you a picture of a ferret [drawing]' and then [ArtInsight] can do the rest and I don't have to keep bugging [her].} --- \textbf{P1}
\end{quote}

P3 wanted to save the final set of descriptions (updated after the audio recording) to peruse after the interaction with her daughter. She already saves pictures of artwork to her phone with metadata, and commonly revisits those, so she envisioned doing the same with artwork analyzed on ArtInsight.

\textbf{Expectations from AI.} Our study sessions revealed occasional mismatches between what children (particularly younger children) expected from AI descriptions \textit{vs.} what the adult BLV family members expected from AI descriptions. Children, especially younger children, expected AI descriptions to accurately explain all parts of their artwork---colors, animals or objects, specific numbers of items on the page, and more. For example, in relation to ArtInsight's initial AI description of one of her drawings (Figure \ref{P3Daughter}), P3's daughter (age 6) picked up on multiple minor, incorrect statements:

\begin{quote}
    \textit{It [the description] said... purple but it's actually blue... it said bunnies, but they're doggies... It said that there was flowers in the basket, but it's bread...} --- \textbf{P3's daughter}
\end{quote}

P2's son (age 7), when responding to ArtInsight's initial AI description of his artwork that commented on parts of the animal being \textit{``outlined''} (Figure \ref{DogScribble}), said: \textit{``I did not outline it!''} To him, there was not a differentiation between the description using \textit{``outlined''} as an artistic descriptive word \textit{vs.} an intended act on his part.

In contrast, BLV family members did not have concerns with AI descriptions having some inaccuracies. P3's mother expressed wanting to hear detailed AI descriptions \textit{``even if it's wrong... because it still told me there is something there.''} P5 explained that she \textit{``already know[s] not to trust these things [AI],''} and she would use instances where she suspected mistakes in an AI description as an opportunity to further probe the AI agent or her granddaughter for more details.

\textbf{Desired additions to the ArtInsight experience.}
P1, P4, and P5 all expressed wanting to prompt the AI agent for more details upon receiving a description, similar to how Be My AI allows users to ask follow-up questions to human or AI agents. While these participants placed emphasis on dialogue with their children, they also wanted the ability to follow up with the AI agent directly.

\section{Case Study with a Blind Family Therapist}
We suspect that ArtInsight might be useful in artwork interpretation contexts beyond mixed visual-ability families. One such context is child therapy \cite{Bosgraaf2020}. To that end, as a complement to our user study with families, we also conducted a qualitative case study with a blind family therapist who uses artwork in her practice. The therapist, T1, is a legally blind woman aged 25-34 who regularly works with children 4-17 years old.

\subsection{Method} 
We followed the same procedure as the study with families for our session with T1, with two main differences: (1) we used two example art pieces sourced from the lead researcher's family (Artwork1 and Artwork2), and (2) we began our session with a semi-structured interview to learn more about T1's experiences as a family therapist. Additionally, the lead researcher acted as the child artist for the evaluation of the audio recording with AI re-prompting component. We compensated T1 \$40/hr. for a 90 minute in-person session. We recorded, transcribed, and analyzed the case study session using the same inductive and deductive coding procedures \cite{Braune_And_Clarke} to arrive at an initial codebook, and peer debriefing \cite{PeerDebriefing} for codebook iteration.

\subsection{Findings}
The findings from our case study highlight how T1 engages with her young patients. Here, we present T1's current practices and challenges as a blind family therapist, as well as T1's evaluation of ArtInsight and subsequent feedback.

\textbf{Current practices and challenges as a blind family therapist.}
T1 began the session by describing how she has used art for child therapy sessions, and art's importance in her work: 

\begin{quote}
    \textit{Art is a backbone aspect of [child] therapy... it's all about expressing yourself and expressing your emotions, but not all children have the vocabulary to express what's going on... they use the drawing to tell a story.}  --- \textbf{T1}
\end{quote}

Typically, T1 relies on sighted colleagues when working with children, as using artwork to express emotions is a common therapy exercise. As an associate therapist, T1 always had a sighted supervisor in her sessions with whom she could consult. But as an independent practitioner, she does not have that same access to sighted help:

\begin{quote}
    \textit{Now, I'm a licensed person, so I don't have anyone to supervise [my sessions]... [in my] new job, though I have the opportunity to work with children, mostly I'm going to try to work with adults... with adults, you don't need to see as much artwork... but if [ArtInsight] is useful for me, maybe I can use it to work with children.} --- \textbf{T1}
\end{quote}

T1 has also trained herself to pick up on extra \textit{``paraverbal cues''} when children describe their artwork to her: 

\begin{quote}
    \textit{I can't see, so... I have to really analyze from a therapist's point of view... I can hear their volume, I can hear their cadence, I can hear their tone. I have to do a lot of gleaning from all those paraverbal cues.} --- \textbf{T1}
\end{quote}

\textbf{Evaluation of ArtInsight.}
Similar to the user study, this part began with a comparison of artwork descriptions by ArtInsight and Be My AI, followed by an evaluation of the additional ArtInsight components. T1 rated \textit{usefulness} equally for both ArtInsight and Be My AI's descriptions on each example artwork (4 for Artwork1, 3 for Artwork2). T1 commented that to rate a description a 5/5 for \textit{Usefulness}, she would expect the description to \textit{``describe the feelings or the emotions on the people's faces.''} For \textit{Perceived Accuracy}, T1 rated the ArtInsight description higher (5) than Be My AI (3) for Artwork1, but for Artwork2, she rated them the same (3). T1 explained that ArtInsight scored higher on \textit{Perceived Accuracy} for Artwork1 because it did a better job of recognizing the child's name on the artwork. Generally, T1 found ArtInsight's initial descriptions to be too long for usefulness within and in-between her sessions:

\begin{quote}
    \textit{[The length] doesn't add value to what I'm trying to get from [the description]... I would pretty much use it after [the session ends]. But even then, we have literally a seven minute break in between sessions.} --- \textbf{T1}
\end{quote}

This concern with length reappeared as T1 explored ArtInsight's "creative" toggle, though T1 did say the creative output told \textit{``more of a story.''} T1 also had feedback for the description language:

\begin{quote}
    \textit{I want to be the one to interpret it, that it's `bold' and `vibrant,' from my own imagination, versus the interpreter doing that interpretation for me. I literally want [AI] to be my eyes, not to be my brain as well.} --- \textbf{T1}
\end{quote}

T1's favorite component was the audio recording and subsequent description regeneration. As the lead researcher played the role of the child artist for T1 to test the audio recording, T1 (unprompted) began asking questions such as: \textit{``Do you and your family go apple picking?''} (in response to the lead researcher describing an apple tree in an example artwork); and \textit{``Who is the person with the pink hair? Is it you?''} (in response to the lead researcher describing a figure in the art). These questions highlighted how, for T1, the raw description by the child is not enough context about an artwork---she also wants to capture the responses to probing questions.

T1 is used to thinking of questions herself about the child's artwork, so she did not find value in the AI-generated questions tab. However, she commented that the questions tab could be good for parents (both sighted and BLV), as many of the parents she works with \textit{``really struggle... to talk to [their] child.''}

Finally, when thinking holistically about when and how she would like to use specific ArtInsight components, T1 explained her preference to not use her phone when she is with children. She expressed wanting to use ArtInsight to get the full artwork description \textit{after} the conclusion of a therapy session, but would need to attain the child's audio recording describing their work \textit{during} the session:

\begin{quote}
    \textit{I'd probably get the audio description first. And then I would take a picture of the final [art] piece. And then have to marry the two [after the session]. And then get the creative [interpretation].} --- \textbf{T1}
\end{quote}

\textbf{Feedback for AI-powered child artwork interpretation systems.} 
T1 commented on three elements that have acted as past deterrents for her using an AI vision application (\textit{e.g.},  Be My AI): (1) the value of time in sessions, (2) the momentary nature of access to children's artwork, and (3) the ease of uploading photos. We also discussed privacy considerations for uploading artwork created during therapy sessions to AI platforms.

With respect to the first deterrent, T1 discussed how the lag between taking a photo of an artwork and receiving an AI description makes it challenging for her to use AI image description systems during her therapy sessions: \textit{``During that [lag] time, children could be doing anything... I need to give them my full attention.''}

T1 also described how access to children's artwork is not always possible past the duration of the sessions, as children frequently take their artwork home. While T1 does try to photograph each art piece, systems such as Be My AI do not have an easy in-application photo upload experience, which acts as a barrier for T1 to use AI-powered description services.

We also discussed privacy implications of uploading artwork created by children in therapy sessions to AI platforms. T1 commented on the steps she takes concerning privacy: 

\begin{quote}
    \textit{It's super anonymous... [In the artwork,] there's no identifying information to the patient. In HIPAA, we're allowed to talk about our patients... as long as we don't identify them. [In the past] I've [used] my supervisor to look [for identifiers] because I had that resource. And usually I ask children to write their name in the back of the artwork. So that's how I protect them.} --- \textbf{T1}
\end{quote}

\section{Discussion}
We introduced and studied ArtInsight, a novel AI-powered mobile prototype for interpreting child-created artwork in mixed visual-ability families. Below, we reflect on our findings from two evaluations with mixed visual-ability families and a blind family therapist and contextualize them in prior work, enumerate considerations for future AI-powered artwork description technology for mixed visual-ability interactions, and reflect on limitations. We interweave opportunities for future work throughout.

\subsection{Reflecting on Key Findings}

Our studies began with a comparison of ArtInsight's initial descriptions to descriptions from Be My AI, a common BLV image description tool. Participants rated \textit{Usefulness} of ArtInsight descriptions higher than Be My AI, appreciating the detailed nature, structure, and artistic language of ArtInsight descriptions. We thus encourage future work for commonly-used BLV AI tools \cite{be_my_ai_2024, seeing_ai} to consider integrating AI prompts specific to interpreting children's artwork.

Of the additional ArtInsight components, BLV adults across our studies found the least value in the AI-generated questions feature, as asking children about their artwork is already common practice. In contrast, \textit{ContextQ} \cite{Dietz_ContextQ} as one of several prior works \cite{Zhang_StoryBuddy, Lin_FishScales} employing AI-generated questions but for co-reading, found high question utilization by parents with their children. Our contrasting results are, perhaps, due to technical implementation differences, distinct contexts (co-reading \textit{vs.} artwork interpretation), or different user populations (\textit{e.g.,} mixed visual-ability families). We encourage further research to investigate the role of AI-generated questions for family dialogue across contexts.

Most BLV family members and children appreciated the creative descriptions component, as it helped bring the artwork \textit{``to life.''} These results highlight the benefits of multiple description variations, extending work by \citet{Li_UnderstandingVisualArtsExperiences} who encouraged supporting "objective" \textit{vs.} "subjective" toggles for visual art descriptions. However, some BLV adults (P4, T1) found the creative description too verbose. T1 additionally discussed wanting a description that interpreted \textit{emotions} in children's artwork, highlighting a use case for description variations beyond family settings. Future research should investigate enabling AI description options beyond "descriptive" and "creative" (such as "verbose" and "succinct"; or "emotional" and "stoic"), as well as allowing for more granular control of the "personality" of the AI description. For example, consider a spectrum from "descriptive" to "creative" in place of a binary toggle, akin to setting the temperature on a generative AI model\footnote{\href{https://learn.microsoft.com/en-us/ai-builder/prompt-modelsettings}{https://learn.microsoft.com/en-us/ai-builder/prompt-modelsettings}}.

Three of the five BLV family members stated the audio recording with AI re-prompting as their favorite component. The ability to capture their children's own interpretation of the work was critically important, supporting the value of the child's narrative \cite{chhedakothary2024}, extending the benefit of child corrections to AI \cite{darth_vader} to include describing child-created artwork, and ultimately reinforcing the work of \citet{bennett_itscomplicated} that highlights the importance of \textit{human agency} in AI-driven descriptions for BLV people.

\subsection{Considerations for Accessible AI-Powered Understanding of Children's Art}
We additionally enumerate high-level considerations for systems supporting AI-powered child artwork understanding for mixed visual-ability groups, extending guidelines proposed by \citet{chhedakothary2024} and highlighting ethical considerations of working with AI systems that use children's data.

\textbf{Balancing accuracy and guesswork in AI across BLV family members and children.} While children prioritize accuracy in AI descriptions, BLV adults are more accepting of minor inaccuracies as long as the description is detailed enough to spark conversation. Notably, both children and BLV adults appreciate when AI makes an informed guess about the artwork (\textit{e.g.}, ``it looks like a bear'') rather than using detailed yet vague terms such as \textit{``a furry animal with round ears and a short tail.''} AI tools should account for the varying needs of BLV adults, sighted children, and both together as different combinations of "users", balancing accuracy with thoughtful guesses to meet expectations.

\textbf{Flexible, independent use of system components.} While we designed ArtInsight to support flexible usage (\textit{i.e.}, before, during, or after interacting with the child around their artwork, DG3), our findings highlight a need for \textit{even greater} control of individual components. For example, T1 wanted a workflow of first audio-recording the child describing their artwork, then at a later stage taking a picture of the artwork, and eventually marrying the two. Related, P1 described her daughter sending her the artwork along with a text description of its context, in which case any initial AI description should accommodate both inputs simultaneously. AI tools should allow flexible, independent use of their components to support diverse mixed visual-ability interaction needs.

\textbf{Supporting the "personality" of AI descriptions desired by the individual(s).} Our findings reveal that BLV individuals and their children value varied description styles. For example, P1 and her 17-year-old daughter, an advanced artist, preferred artistic language. T1 wanted AI descriptions to convey the emotions of people in children's drawings. P4 and his son had contrasting preferences---P4 preferred the descriptive while his son preferred the creative. AI tools should carefully balance these needs in their design, allowing for BLV family members and their children to individually or collectively configure their desired AI personalities.

\textbf{Ethical implications of using AI systems with children's data.} In utilizing AI systems to interpret child-created artifacts, ethical considerations around privacy and security are paramount. To address these concerns in our research, we only uploaded child-created artworks to our GPT backend with explicit parental consent. Families could opt out of sharing art, and we anonymized any shared artwork by obfuscating identifiable markers such as children's names when taking the initial photo of the art. Furthermore, any voice recordings of children were stored locally on the researcher's mobile device, and only transcripts were uploaded to the backend. We encourage similar future AI systems to implement these measures and more to protect the privacy of children and their families, such as following an "opt-in" method of using uploaded data for additional model training and promoting AI literacy among users to foster data awareness. Moreover, future systems could run local-only models on the phone without cloud requirements and attempt to obfuscate identifiable markers automatically.

\subsection{Limitations}
Our study had three key limitations: a small participant pool, the use of a tool designed for a different demographic in the family therapist case study, and potential model bias in the technical evaluation process. First, we conducted the user study as a lab study with a limited number of family groups (four parents and one grandparent with their respective children), which may affect the generalizability of our findings. A study with more families and diverse BLV family member-sighted child relationships could uncover further insights. Another study format such as a longitudinal study could also reveal how families can organically use ArtInsight.

Second, we included a case study with a BLV family therapist to offer a new perspective on the potential benefits of ArtInsight outside of families. However, the system was primarily designed for BLV family members and their sighted children, which may have affected our case study. We encourage future work in supporting BLV family therapists to work independently with children.

Lastly, our AI model evaluation experiments resulted in a GPT model performing the best, but there is a potential bias in the evaluation process, as the GPT-based LLM Scorer used for evaluation may have a natural alignment with its own outputs. While we used human spot-checking as a possible mitigation, future work should incorporate a more diverse set of scoring frameworks to minimize bias and ensure a fairer comparison across models.

\section{Conclusion}
In this work, we present technical and user study evaluations of \textit{ArtInsight}, a novel AI-powered artwork interpretation system to facilitate deeper engagement with child-created artwork in mixed visual-ability families. We also introduce a rubric to determine the quality of AI descriptions of child-created artwork. Our user studies reveal ArtInsight descriptions are more useful for participants compared to descriptions from a popular BLV AI tool, \textit{Be My AI}. Participants additionally valued ArtInsight's novel features, including the creative descriptions and audio recordings with AI re-prompting, as these fostered greater connection to the artwork and centered the narrative of the child artists. Our work ultimately has implications for Human-AI researchers, those working in nuanced use cases of multimodal LLMs, and accessibility practitioners researching mixed-ability interactions.

\begin{acks}
We thank Anukriti Kumar, Jennifer Mankoff, and Aashaka Desai for their help. This work was supported in part by the National Science Foundation grant \#2125087 and The Mani Charitable Foundation. Any opinions, findings, conclusions or recommendations expressed in our work are those of the authors and do not necessarily reflect those of any supporter.
\end{acks}

\bibliographystyle{ACM-Reference-Format}
\bibliography{references.bib}

\clearpage
\onecolumn

\appendix \def\thesection{\arabic{section}}
 \setcounter{section}{0}
\section{APPENDIX}
\subsection{Final Prompt for ArtInsight Initial AI Description}
\label{FinalPrompt}
\begin{quote}
    Generate a descriptive description of the artwork in paragraph form (no bullets or numbered points). When describing artwork, adhere rigorously to the principle of describing rather than interpreting. Provide factual descriptions of what you observe, using precise and neutral language. Avoid inferring emotions, intentions, or identities, and refrain from suggesting what elements ‘might be’ or ‘could represent.’ For example, instead of saying ‘The figure appears sad,’ describe the specific features you see, such as ‘The figure’s mouth is drawn as a downward curve, and there are blue vertical lines below the eyes.’ Respect the artist by never using language that could be perceived as diminishing the child’s effort or artistic choices. Avoid terms like ‘simple,’ ‘rough,’ ‘messy,’ or ‘childish.’ Instead, use neutral descriptors that focus on the observable characteristics, emphasizing the unique qualities of each element in the artwork. Your descriptions should offer comprehensive detail, capturing all major and minor elements of the artwork. Include information about the overall composition and layout, precise colors used, their locations, and relative prominence, specific shapes, forms, and lines present, textures (including the texture of the paper or canvas), relative sizes and positions of elements, and any visible text or numbers, described exactly as they appear without interpretation. Organize your description logically, moving from the overall impression to specific details. Use clear, concise language that a blind parent can easily visualize. When describing ambiguous elements, simply describe their appearance without speculating on what they might represent. Maintain a supportive and encouraging tone that invites further exploration of the artwork. Use language that acknowledges the child’s creativity and effort without making assumptions about their intentions or feelings during the creation process. Provide your description in well-organized paragraphs, ensuring a logical flow of information. Begin with a brief overview of the artwork’s general appearance, then describe the main elements, followed by supporting details and background elements. Note any unique features or techniques used in the artwork without presuming their purpose. Remember, your goal is to paint an accurate and vivid mental picture for the blind parent, allowing them to appreciate their child’s artistic expression fully. Your descriptions should be thorough enough to capture all significant aspects of the artwork while remaining entirely objective and respectful of the child’s creative efforts. Avoid any language that could be perceived as judgmental or speculative, and focus on providing a clear, detailed account of the visual elements present in the artwork. Do not ask questions or suggest interpretations in your descriptions. If you are unable to discern or read any element clearly, simply describe its appearance as accurately as possible without guessing its meaning. Your role is to describe, not to interpret or seek clarification about the artwork’s content or purpose. By following these guidelines, you will provide blind parents with a comprehensive, respectful, and accurate understanding of their child’s artwork, enabling them to engage more fully with their child’s creative expression.
\end{quote}

\subsection{Additional Prompt Instructions for Creative Description}
\begin{quote}
    Generate a creative description of the artwork in paragraph form (no bullets or numbered points). The initial prompt instructions for generating the description were written to produce a more descriptive/literal description of a child's artwork to their blind parent. Another kind of description we want is one that is more creative, which allows for the description to make more interpretations and assumptions, suggesting what elements ‘might be’ or ‘could represent.’ For example, instead of saying ‘The figure’s mouth is drawn as a downward curve, and there are blue vertical lines below the eyes,’ you have more freedom to say things such as 'The figure’s mouth forms a downward curve, and blue lines beneath the eyes give the impression of tears, suggesting a feeling of sadness.' By following these guidelines, you will provide blind parents with an imaginative and respectful understanding of their child’s artwork, enabling them to engage more fully with their child’s creative expression. Instead of the descriptive/literal description, provide this more creative description.
\end{quote}

\subsection{Additional Prompt Instructions for Generated Questions}

\begin{quote}
Generate 3 questions the parent can ask the child about their artwork.
\end{quote}

\newpage
\subsection{Example LLM Scorer Scores and Rationale}
\begin{table*}[h]
  \centering
  \renewcommand{\arraystretch}{1.1}
  \resizebox{\textwidth}{!}
  {
    \begin{tabular}{p{0.07\linewidth} p{0.31\linewidth} p{0.31\linewidth} p{0.31\linewidth}}
      \textbf{}
       & \begin{center}
          \includegraphics[width=\linewidth]{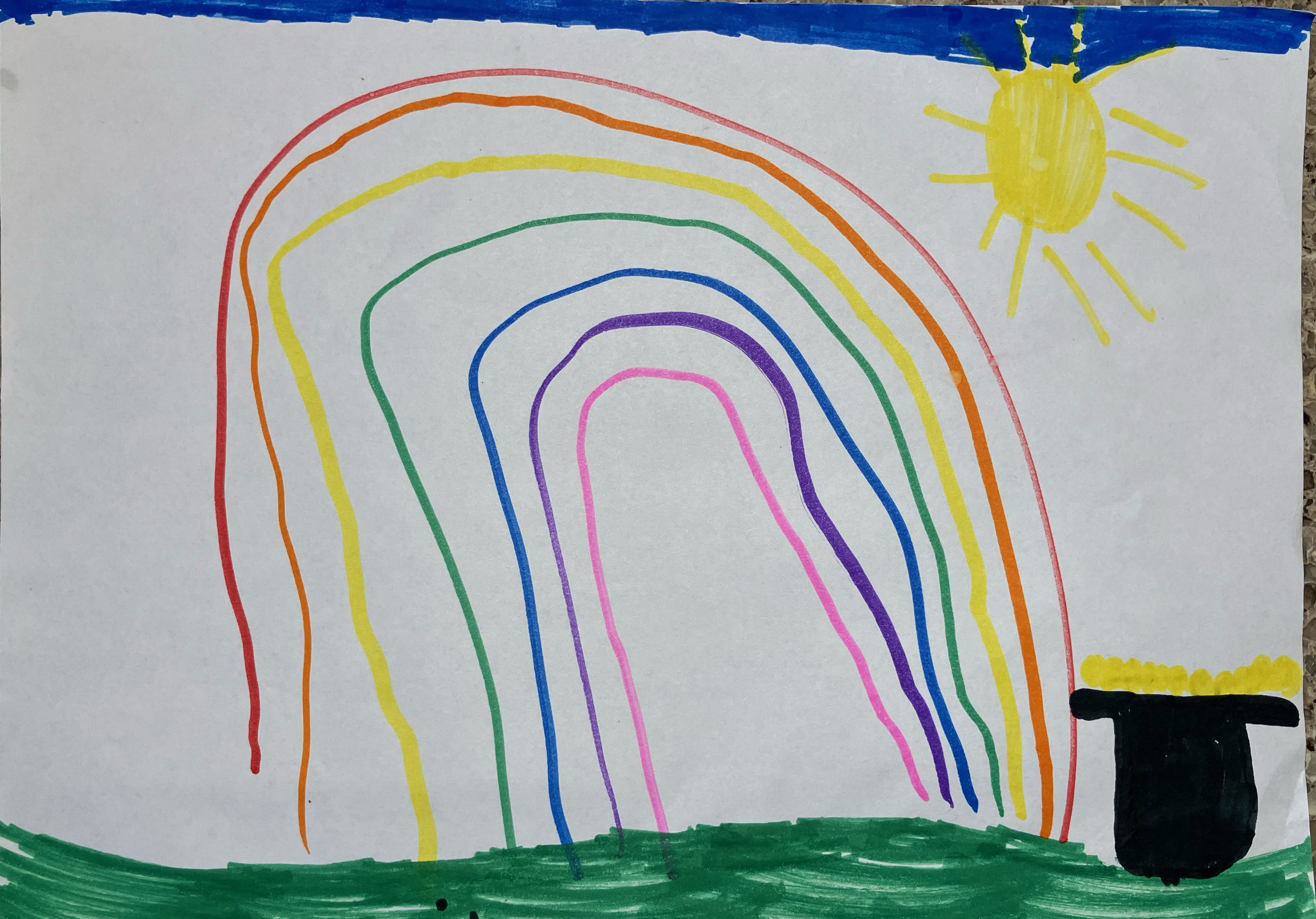}
          \textbf{\textit{Rainbow}}
        \end{center}%
        & \begin{center}
          \includegraphics[width=\linewidth]{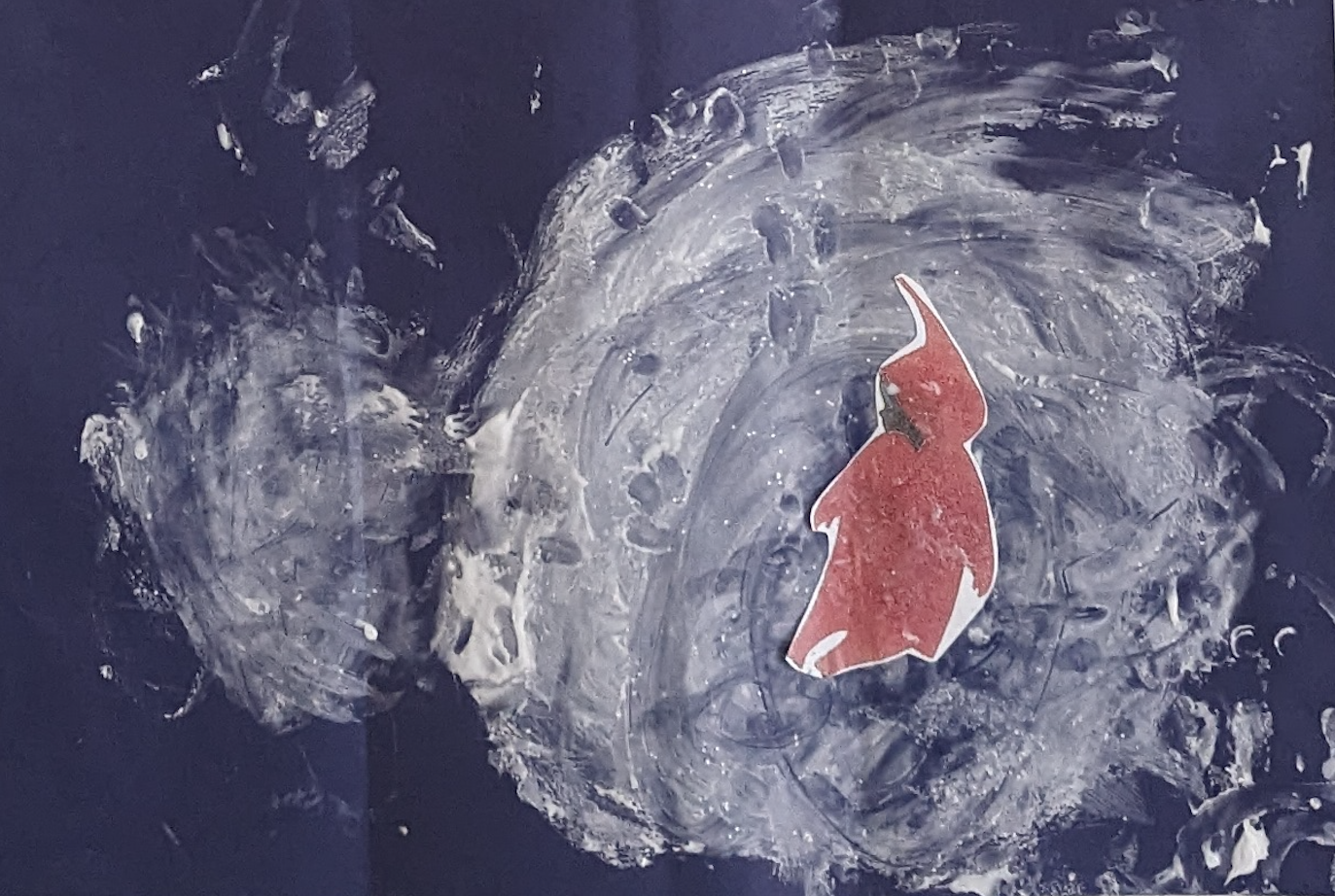}
          \textbf{\textit{AbstractBlueWhiteRed}}
          \end{center}%
        & \begin{center}
          \includegraphics[width=\linewidth]{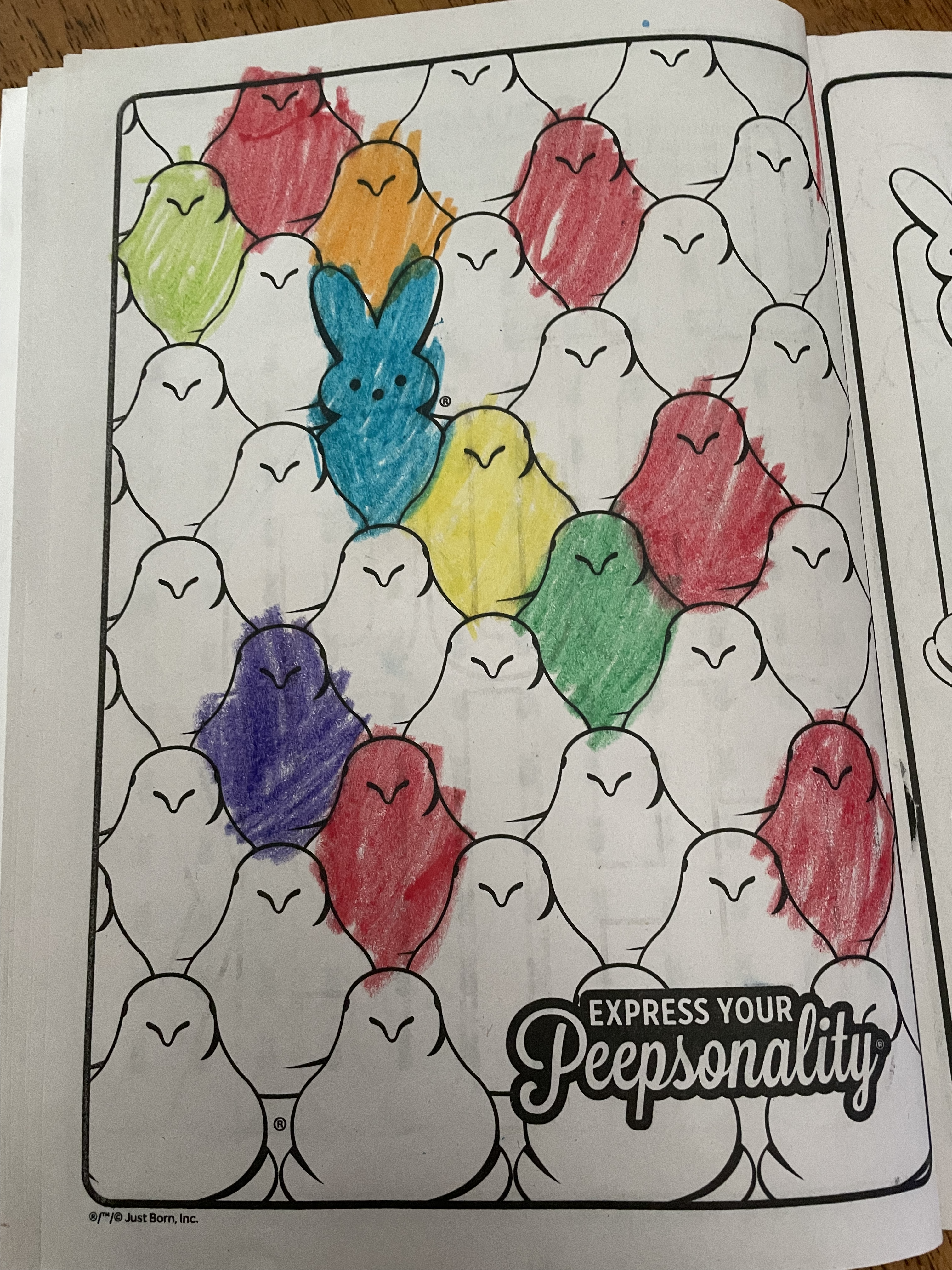}
          \textbf{\textit{Peeps}}
          \end{center}%
        \\
      \hline
        \textbf{Claude 3.5 Sonnet}
        &
        \begin{tabular}{| p{0.17\linewidth} | p{0.16\linewidth} | p{0.17\linewidth} | p{0.17\linewidth} |}
          \textbf{R.A} & \textbf{R.B} & \textbf{R.C} & \textbf{R.D} \\
          \hline
          3 & 4 & 4 & 4 \\
          \hline
        \end{tabular}
        \vspace{0.2em}
        \newline
        Makes assumptions, such as the black object being a "pot or cauldron" and the yellow line representing "gold or something valuable inside."
        & 
        \begin{tabular}{| p{0.17\linewidth} | p{0.16\linewidth} | p{0.17\linewidth} | p{0.17\linewidth} |}
          \textbf{R.A} & \textbf{R.B} & \textbf{R.C} & \textbf{R.D} \\
          \hline
          2 & 4 & 4 & 4 \\
          \hline
        \end{tabular}
        \vspace{0.2em}
        \newline
        Assumptions include the red figure being a "person or creature" and suggesting themes like "isolation, adventure, or standing out in a chaotic world." 
        &
        \begin{tabular}{| p{0.17\linewidth} | p{0.16\linewidth} | p{0.17\linewidth} | p{0.17\linewidth} |}
          \textbf{R.A} & \textbf{R.B} & \textbf{R.C} & \textbf{R.D} \\
          \hline
          2 & 4 & 4 & 3 \\
          \hline
        \end{tabular}
        \newline
        Assumes blue shape may be a "gingerbread man or a little person." Assumes purpose of the text. Should specify the placement of colors. \hfill

        \textit{-1 Misc}: Misidentifies the blue shape.
        \\
        \hline
        \textbf{Score}
        &
        \textbf{15/16}
        & 
        \textbf{14/16}
        &
        \textbf{12/16}
        \\
        \hline
        \textbf{GPT-4 Turbo}
        &
        \begin{tabular}{| p{0.17\linewidth} | p{0.16\linewidth} | p{0.17\linewidth} | p{0.17\linewidth} |}
          \textbf{R.A} & \textbf{R.B} & \textbf{R.C} & \textbf{R.D} \\
          \hline
          3 & 4 & 4 & 4 \\
          \hline
        \end{tabular}
        \newline
        Makes an assumption about child intent: the black pot with a yellow band is "the classic tale of a pot of gold at the end of the rainbow."
        & 
        \begin{tabular}{| p{0.17\linewidth} | p{0.16\linewidth} | p{0.17\linewidth} | p{0.17\linewidth} |}
          \textbf{R.A} & \textbf{R.B} & \textbf{R.C} & \textbf{R.D} \\
          \hline
          2 & 4 & 4 & 4 \\
          \hline
        \end{tabular}
        \newline
        Assumptions about the intent and themes: "a moment of calm... within a stormy or chaotic environment," and "themes such as peace amid chaos..."
        &
        \begin{tabular}{| p{0.17\linewidth} | p{0.16\linewidth} | p{0.17\linewidth} | p{0.17\linewidth} |}
          \textbf{R.A} & \textbf{R.B} & \textbf{R.C} & \textbf{R.D} \\
          \hline
          3 & 4 & 4 & 4 \\
          \hline
        \end{tabular}
        \newline
        Assumes intent, such as: the choice to color one Peep differently is due to "a sense of standing out" and "mirroring a personal expression."
        \\
        \hline
        \textbf{Score}
        &
        \textbf{15/16}
        & 
        \textbf{14/16}
        &
        \textbf{15/16}
        \\
        \hline
        \textbf{GPT-4o}
        &
        \begin{tabular}{| p{0.17\linewidth} | p{0.16\linewidth} | p{0.17\linewidth} | p{0.17\linewidth} |}
          \textbf{R.A} & \textbf{R.B} & \textbf{R.C} & \textbf{R.D} \\
          \hline
          4 & 4 & 4 & 4 \\
          \hline
        \end{tabular}
        \newline
        Respectful and celebrates the child's creativity with positive language like "joyful" and "vibrant", is detailed and covers colors and brushstrokes, and captures all major elements.
        & 
       \begin{tabular}{| p{0.17\linewidth} | p{0.16\linewidth} | p{0.17\linewidth} | p{0.17\linewidth} |}
          \textbf{R.A} & \textbf{R.B} & \textbf{R.C} & \textbf{R.D} \\
          \hline
          2 & 4 & 4 & 4 \\
          \hline
        \end{tabular}
        \newline
        Suggesting the background is a "night sky or deep space" and that the red figure "might be a character wearing a hood or a cloak." \hfill

        \textit{-1 Misc}: Asking "What does your child say about their creation?"
        &
        \begin{tabular}{| p{0.17\linewidth} | p{0.16\linewidth} | p{0.17\linewidth} | p{0.17\linewidth} |}
          \textbf{R.A} & \textbf{R.B} & \textbf{R.C} & \textbf{R.D} \\
          \hline
          4 & 4 & 4 & 4 \\
          \hline
        \end{tabular}
        \newline
        Avoids making assumptions, is respectful and uses positive language like "fun and eye-catching pattern", is detailed and covers aspects like colors and placement, and captures all major elements.
        \\
        \hline
        \textbf{Score}
        &
        \textbf{16/16}
        & 
        \textbf{13/16}
        &
        \textbf{16/16}
        \\
        \hline
        \textbf{Gemini 1.5 Flash}
        &
        \begin{tabular}{| p{0.17\linewidth} | p{0.16\linewidth} | p{0.17\linewidth} | p{0.17\linewidth} |}
          \textbf{R.A} & \textbf{R.B} & \textbf{R.C} & \textbf{R.D} \\
          \hline
          2 & 4 & 3 & 4 \\
          \hline
        \end{tabular}
        \newline
        Too simple, makes assumptions about the intent behind the artwork ("fun day"), asks speculative questions. \hfill

        \textit{-1 Misc}: Asking "What do you think it represents to your child?".
        & 
        \begin{tabular}{| p{0.17\linewidth} | p{0.16\linewidth} | p{0.17\linewidth} | p{0.17\linewidth} |}
          \textbf{R.A} & \textbf{R.B} & \textbf{R.C} & \textbf{R.D} \\
          \hline
          1 & 3 & 2 & 3 \\
          \hline
        \end{tabular}
        \newline
        Assumptions about the white areas "look like clouds", the scene is "in the sky". Saying "maybe this is a scene in the sky?" could be minimizing to the child's effort. Also too simple and not all major elements captured. \hfill

        \textit{-1 Misc}: Asks "What do you think?"
        &
        \begin{tabular}{| p{0.17\linewidth} | p{0.16\linewidth} | p{0.17\linewidth} | p{0.17\linewidth} |}
          \textbf{R.A} & \textbf{R.B} & \textbf{R.C} & \textbf{R.D} \\
          \hline
          4 & 4 & 3 & 3 \\
          \hline
        \end{tabular}
        \newline
        Could provide more detail about the specific arrangement and the overall pattern of the figures. Also misses identifying the text, "Express your Peepsonality".
        \\
        \hline
        \textbf{Score}
        &
        \textbf{12/16}
        & 
        \textbf{8/16}
        &
        \textbf{14/16}
        \\
        \hline
    \end{tabular}
  }

  \caption{Three example images from our dataset with the LLM Scorer scores for each Rubric guideline (R.A---presumptive, R.B---reductive, R.C---too simple, R.D---all elements captured), the reasoning for points lost, and the total scores.}
  \label{MasterTable}  \vspace{-1.4em}
\end{table*}

\newpage
\subsection{LLM Scorer Results for Model Comparison}

\begin{table*}[h]
\centering
\small 
\setlength{\extrarowheight}{2pt} 
\resizebox{0.85\textwidth}{!}{ 
\begin{tabular}{lllll}
\toprule 
\rowcolor{white}  
\textbf{Image ID} & \textbf{Claude 3.5 Sonnet} & \textbf{GPT-4 Turbo} & \textbf{GPT-4o} & \textbf{Gemini 1.5 Flash} \\
\midrule

1 & 16 & 13 & 16 & 12 \\
2 & 14 & 15 & 16 & 9 \\
3 & 11 & 14 & 14 & 5 \\
4 & 13 & 14 & 16 & 12 \\
5 & 15 & 16 & 16 & 12 \\
6 & 15 & 16 & 16 & 12 \\
7 & 16 & 15 & 16 & 12 \\
8 & 16 & 15 & 16 & 13 \\
9 & 15 & 15 & 15 & 8 \\
10 & 16 & 14 & 16 & 8 \\
11 & 15 & 14 & 16 & 13 \\
12 & 16 & 14 & 16 & 14 \\
13 & 13 & 13 & 16 & 11 \\
14 & 16 & 13 & 16 & 12 \\
15 & 15 & 15 & 16 & 15 \\
16 (\textit{Rainbow}) & 15 & 15 & 16 & 12 \\
17 & 14 & 16 & 16 & 12 \\
18 & 14 & 15 & 16 & 10 \\
19 & 15 & 15 & 16 & 12 \\
20 & 16 & 16 & 16 & 13 \\
21 & 16 & 15 & 16 & 10 \\
22 (\textit{AbstractBlueWhiteRed}) & 14 & 14 & 13 & 8 \\
23 & 15 & 16 & 16 & 11 \\
24 & 15 & 16 & 15 & 13 \\
25 (\textit{Peeps}) & 12 & 15 & 16 & 14 \\
26 & 16 & 12 & 16 & 14 \\
27 & 16 & 14 & 16 & 11 \\
28 & 16 & 16 & 16 & 12 \\
29 & 15 & 15 & 16 & 12 \\
30 & 15 & 15 & 15 & 13 \\
Average & 14.87 & 14.7 & 15.73 & 11.5 \\ 

\bottomrule
\end{tabular}
}
\vspace{0.5em}
\caption{The scores of how different models performed across 30 images of children's artwork using our v0 prompt, with the maximum score possible being 16. The three artworks highlighted with more score details in Table \ref{MasterTable} are identified inline.}
\label{Test}
\vspace{-1em}
\end{table*}

\newpage
\subsection{LLM Scorer and Human Scorer Evaluation of ArtInsight \textit{vs.} Be My AI}
\label{ArtInsightBeMyAIScores}\enlargethispage{30pt}

\begin{table*}[h]
\centering
\LARGE 
\setlength{\extrarowheight}{5pt} 
\resizebox{\textwidth}{!}{ 
\begin{tabular}{cclcccc}
\toprule 
\rowcolor{white}  
\textbf{PID} & \textbf{Image} & \textbf{Application} & \textbf{LLM Scorer} & \textbf{Researcher 1} & \textbf{Researcher 2} & \textbf{External AI Researcher} \\
\midrule

1 & 1 & Be My AI & 13 & 11 & 13 & 11 \\
1 & 1 & ArtInsight & 16 & 16 & 15 & 14 \\
1 & 2 & Be My AI & 3 & 11 & 13 & 11 \\
1 & 2 & ArtInsight & 16 & 16 & 14 & 14 \\

2 & 1 & Be My AI & 14 & 12 & 11 & 14 \\
2 & 1 & ArtInsight & 16 & 15 & 14 & 15 \\
2 & 2 & Be My AI & 6 & 10 & 12 & 10 \\
2 & 2 & ArtInsight & 16 & 13 & 15 & 13 \\

3 & 1 & Be My AI & 12 & 10 & 11 & 7 \\
3 & 1 & ArtInsight & 16 & 14 & 13 & 14 \\
3 & 2 & Be My AI & 14 & 12 & 13 & 13 \\
3 & 2 & ArtInsight & 16 & 13 & 15 & 16 \\

4 & 1 & Be My AI & 14 & 10 & 13 & 13 \\
4 & 1 & ArtInsight & 16 & 13 & 15 & 15 \\
4 & 2 & Be My AI & 14 & 11 & 14 & 5 \\
4 & 2 & ArtInsight & 16 & 16 & 16 & 16 \\
4 & 3 & Be My AI & 14 & 10 & 11 & 8 \\
4 & 3 & ArtInsight & 15 & 13 & 12 & 14 \\

5 & 1 & Be My AI & 14 & 10 & 10 & 3 \\
5 & 1 & ArtInsight & 16 & 15 & 13 & 13 \\
5 & 2 & Be My AI & 14 & 11 & 9 & 10 \\
5 & 2 & ArtInsight & 16 & 14 & 13 & 12 \\

\bottomrule
\end{tabular}
}
\vspace{0.5em}
\caption{Automated and human scoring of the initial AI descriptions provided by Be My AI and ArtInsight for each image explored by participants. Scores are out of 16 points per our rubric.}
\label{Scoring}
\vspace{-1em}
\end{table*}

\end{document}